\documentclass[12pt]{article}
\usepackage{amsthm}
\usepackage{eurosym}
\usepackage{amsfonts}
\usepackage{amssymb,amsmath}
\usepackage[dvips]{color}
\usepackage{amscd}
\usepackage{tikz}
\usepackage{mathtools}
\usepackage{graphicx}
\usepackage{caption}
\usepackage{subcaption}
\usepackage[all,cmtip]{xy}

\newcommand{\xdownarrow}[1]{ {\left\downarrow\vbox to #1{}\right.\kern-\nulldelimiterspace} }
\usetikzlibrary{decorations.pathreplacing}

\usetikzlibrary{patterns}

\def\and{\mathrm{and}}

\newtheorem{prop}{Proposition}

\newcommand{\ee}{\end{equation}}
\newcommand{\bea}{\begin{eqnarray}}
\newcommand{\eea}{\end{eqnarray}}
\newcommand{\beas}{\begin{eqnarray*}}
\newcommand{\eeas}{\end{eqnarray*}}
\newcommand{\ba}{\begin{array}}
\newcommand{\ea}{\end{array}}

\newcommand{\nbox}{{\,\lower0.9pt\vbox{\hrule \hbox{\vrule height 0.2 cm \hskip 0.19 cm \vrule height 0.2 cm}\hrule}\,}}

\def\href#1#2{#2}
\theoremstyle{plain}

\input{tcilatex}

\begin{document}

\begin{titlepage}
\hfill
\vbox{
    \halign{#\hfil         \cr
           } 
      }  

\hbox to \hsize{{}\hss \vtop{ \hbox{}

}}

%

\vspace*{20mm}

\begin{center}

{\large \textbf{Coherent state superpositions, entanglement and \\ \vspace{0.4cm} gauge/gravity correspondence}}

{\normalsize \vspace{10mm} }

{\normalsize {Hai Lin${}^{1,2}$}, Yuwei Zhu${}^{1,2}$}

{\normalsize \vspace{10mm} }

{\small \emph{${}^1$\textit{Yau Mathematical Sciences Center, Tsinghua University,
Beijing 100084, China
}} }

{\normalsize \vspace{0.2cm} }

{\small \emph{$^2$\textit{Department of Mathematical Sciences, Tsinghua University,
Beijing 100084, China
\\
}} }

{\normalsize \vspace{0.4cm} }

\end{center}

\begin{abstract}

We focus on two types of coherent states, the coherent states of multi graviton states and the coherent states of giant graviton states, in the context of gauge/gravity correspondence. We conveniently use a phase shift operator and its actions on the superpositions of these coherent states. We find $N$-state Schrodinger cat states which approach the one-row Young tableau states, with fidelity between them asymptotically reaches 1 at large $N$. The quantum Fisher information of these states is proportional to the variance of the excitation energy of the underlying states, and characterizes the localizability of the states in the angular direction in the phase space. We analyze the correlation and entanglement between gravitational degrees of freedom using different regions of the phase space plane in bubbling AdS. The correlation between two entangled rings in the phase space plane is related to the area of the annulus between the two rings. We also analyze two types of noisy coherent states, which can be viewed as interpolated states that interpolate between a pure coherent state in the noiseless limit and a maximally mixed state in the large noise limit.

\end{abstract}

\end{titlepage}

\vskip 1cm

\section{Introduction}

\label{sec_introduction} \renewcommand{\theequation}{1.\arabic{equation}} %
\setcounter{equation}{0}

The gauge/gravity correspondence \cite%
{Maldacena:1997re,Gubser:1998bc,Witten:1998qj} is a nontrivial
correspondence between a quantum system without gravity on the boundary and
a quantum theory with gravity in the bulk. It provides a method for studying
quantum gravity by quantum field theory on the boundary of the spacetime.
The correspondence allows us to perform calculations related to superstring
theory and quantum gravity from working on the quantum field theory side. On
the other hand, the superstring theory provides the UV completion of the
supergravity, and is hence a UV-complete quantum gravity. The correspondence
also reveals the nature of the emergent spacetime \cite%
{Horowitz:2006ct,Rangamani:2016dms,Berenstein:2005aa,Koch:2009gq}. The bulk
emerges dynamically from the quantum mechanical description that lives in
fewer dimensions on the boundary.

The boundary system is described by a quantum field theory or quantum
mechanics with a well-defined global time near the boundary. Hence, the
boundary theory has a well defined Hilbert space. Moreover, there is a
correspondence between observables of the bulk spacetime and observables of
the boundary. Progress on the detailed dictionary between the quantum field
theory side and the gravity side, has been made tremendously. For states
that are in the same Hilbert space, we can perform superpositions of these
states. We can also perform other quantum operations \cite{Quantum
information} such as unitary transformations on these states.

In the gauge/gravity correspondence, the quantum field theory side of the
duality is an example of a quantum system with many degrees of freedom. In
such a many-body quantum system, quantum correlations and quantum
entanglement are generic features \cite{Horodecki:2009zz}. Quantum
correlations provide important resources for information processing, and are
important in quantum information theory. There are states that are of
interest both in the gravity side and in quantum information theory, such as
coherent states and their superpositions and entanglement. Moreover, there
are also Young tableau states, which are entangled states.

In the context of this correspondence, there are backreacted geometries that
correspond to highly excited states in the quantum field theory side, such
as the bubbling geometries \cite{Lin:2004nb,Corley:2001zk,Berenstein:2004kk}%
. The states in the Hilbert space of the quantum field theory are explicitly
mapped to the gravity side. Analysis in the field theory side shows that
these different configurations live in the same Hilbert space. Since they
live in the same Hilbert space, one can perform quantum operations allowed
by quantum mechanics, for example one can superpose states and compute
transition probabilities between different states, e.g. \cite%
{Brown:2006zk,Berenstein:2017abm,Diaz:2015tda,Lin:2017dnz}.

We focus on states which have interesting gravitational properties. One
interesting type of states are coherent states \cite{Berenstein:2017abm}.
Gravity dual of coherent states has been analyzed. It has also been shown
that superposition of topologically trivial states can form topologically
nontrivial states in the context of quantum gravity theory \cite%
{Berenstein:2017abm}. In this paper, we focus on two highly interesting
classes of coherent states. One is the coherent state of multi traces, and
another is the coherent state of multi columns of Young tableaux. Another
interesting class of states are Young tableau states. They are also
entangled states in the tensor product of the multi-trace Hilbert spaces
\cite{Berenstein:2017abm,Lin:2017dnz}. These states contain nontrivial
entanglement stored between different multi-trace Hilbert spaces.

The phase space also plays an useful role in identifying the states. The map
between the states in the gravity side and in the field theory side can also
be done by using Wigner function representation of the states in the phase
space, e.g. \cite{Balasubramanian:2005mg,Mandal:2005wv}. We work on
correlations and entanglement between different regions of the phase space
plane \cite{Simon:2018laf,Lin:2017dnz}. These setups provide a laboratory
for studying quantum gravitational questions.

The organization of this paper is as follows. In Section \ref{sec 2}, we
overview a class of single mode and multi mode coherent states, which are
central in the concepts and for the preparation of later discussions. We
define a generalized phase shift operator and parity operator, acting on the
superpositions of coherent states. In Section \ref{sec 3}, we find
Schrodinger cat states, and in particular $N$-state Schrodinger cats which
have a limit approaching Young tableau states. In Section \ref{sec 4}, we
analyze noisy coherent states by adding random traces on top of pure
coherent states. And then in Section \ref{sec 5}, we analyze a different
class of coherent states, which are coherent states of multi columns of
Young tableaux. In Section \ref{sec 6}, we make use of quantum detection
theory to distinguish microstates in the systems discussed in this paper. In
Section \ref{sec 7}, we analyze correlations and entanglement between
different regions of the phase space, and point out that they can be
understood in terms of connectivity between different regions of spacetime.
Finally, we discuss our results and draw some conclusions in Section \ref%
{sec_discussion}.

\section{Class of pure coherent states}

\label{sec 2} \renewcommand{\theequation}{2.\arabic{equation}} %
\setcounter{equation}{0} \renewcommand{\thethm}{2.\arabic{thm}} %
\setcounter{thm}{0} \renewcommand{\theprop}{2.\arabic{prop}} %
\setcounter{prop}{0}

\subsection{Class of single mode and multi mode pure coherent states}

\label{sec 2.1}

In this subsection, we discuss a class of single mode and multi mode
coherent states. This class was constructed in \cite{Berenstein:2017abm},
and analyzed in further details in \cite{Lin:2017dnz}. Many later sections
are closely related to the concept and method in this subsection, hence we
overview them in details for the preparation of later discussions. There is
also a different class of coherent states as coherent states of multi
columns of Young tableaux in Sec. \ref{sec 5}.

Consider the Hilbert space factorizes as $\mathcal{H}$ $=\mathcal{H}%
_{1}\otimes \mathcal{H}_{2}\otimes \dots =\otimes _{k}\mathcal{H}_{k}$. Here
$\mathcal{H}_{k}$ is the Hilbert space for mode $k\in \mathbb{Z}_{>0}$. The
creation and annihilation operators for mode $k$ are $a_{k}^{\dagger }$ and $%
a_{k}$. Their commutation relations are
\begin{equation}
\lbrack a_{k},~a_{k^{\prime }}^{\dagger }]=k\delta _{kk^{\prime }},
\label{commutation_relation_01}
\end{equation}%
with the appropriate normalization convention. The state in mode $k$, with
occupation number $l$, is
\begin{equation}
t_{k}^{l}=(a_{k}^{\dagger })^{l}|0\rangle _{k},
\end{equation}%
where $|0\rangle _{k}~$is the vacuum of $\mathcal{H}_{k}$ and $%
t_{k}^{0}=|0\rangle _{k}$. The state $\frac{1}{\sqrt{l!k^{l}}}%
(a_{k}^{\dagger })^{l}|0\rangle _{k}$ has unit norm.

This construction works generally for systems having a similar Hilbert
space, and in the context of half BPS sector of the gauge theory in which
gauge invariant observables can be constructed from a complex matrix $Y$,
the $t_{k}$ corresponds to $\mathrm{Tr}(Y^{k})$. In that case,%
\begin{equation}
t_{k}=a_{k}^{\dagger }\vert 0\rangle _{k}=\frac{1}{\sqrt{N^{k}}}\text{%
\textrm{Tr}}(Y^{k}),
\end{equation}%
and $t_{k}^{l}=(a_{k}^{\dagger })^{l}\vert 0\rangle _{k}=(\frac{1}{\sqrt{%
N^{k}}}$\textrm{Tr}$(Y^{k}))^{l}$.~The $t_{k}^{l}$ and $t_{k}^{0}$ are the
analogs of the photon states $\vert l\rangle $ and $\vert 0\rangle $ in
quantum optics and quantum information theory. In the context of gauge
theory, the prefactors involving $N$ can be calculated by gauge theory
computations, e.g. \cite%
{Corley:2002mj,Kristjansen:2002bb,Koch:2008cm,Berenstein:2017abm}. In this
paper, we work in the large $N$ limit, because in this limit, the multi
traces provide good orthogonality property. In the context of gauge/string
duality, $t_{k}$ is also a closed string state.

We consider coherent states generalized from the coherent states of photons
\cite{Glauber:1963fi,Zhang:1990fy,Quantum}. A general \emph{multi-mode}
coherent state can be written as, in the language of the creation operator $%
a_{k}^{\dagger }$,
\begin{align}
|Coh(\{\Lambda _{k}\})\rangle & =\prod_{k}\exp (\Lambda _{k}\frac{%
a_{k}^{\dagger }}{k})|0\rangle _{k}  \label{coh_02} \\
& =\prod_{k}(\sum_{l_{k}=0}^{\infty }\frac{1}{l_{k}!}(\Lambda _{k}\frac{%
a_{k}^{\dagger }}{k})^{l_{k}})|0\rangle _{k},
\end{align}%
where $\Lambda _{k}\in \mathbb{C}$. In the language of $t_{k}$,%
\begin{align}
|Coh(\{\Lambda _{k}\})\rangle & =\prod_{k}\exp (\Lambda _{k}\frac{t_{k}}{k})
\label{coh_01} \\
& =\prod_{k}(\sum_{l_{k}=0}^{\infty }\frac{1}{l_{k}!}(\Lambda _{k}\frac{t_{k}%
}{k})^{l_{k}}).
\end{align}%
The $\{\Lambda _{k}\}_{k=1,...,\infty }$ in $Coh(\{\Lambda _{k}\})~$is a
family of complex parameters for the modes $k\in \mathbb{Z}_{>0}$. These
coherent states are at the level of multi-mode coherent states in the
Hilbert space $\otimes _{k}\mathcal{H}_{k}$.

These are \emph{pure} coherent states. They are the eigenstates of the
annihilation operator $a_{k}$ with eigenvalue $\Lambda _{k},$%
\begin{equation}
a_{k}\vert Coh(\{\Lambda _{k}\})\rangle =\Lambda _{k}\vert Coh(\{\Lambda
_{k}\})\rangle .  \label{coherent state condition}
\end{equation}%
We have our convention in the expression of the coherent states (\ref{coh_02}%
).

The single-mode coherent state is a special case. We denote $\vert
Coh(\{\Lambda _{k}\})\rangle $ for the multi-mode one, and $\vert
coh(\Lambda _{k})\rangle _{k}=\exp (\Lambda _{k}\frac{a_{k}^{\dagger }}{k}%
)\vert 0\rangle _{k}$ for the single-mode one in mode $k$. For a single mode
$k$, the pure coherent state, with unit norm, is%
\begin{equation}
\frac{1}{\sqrt{\mathcal{N(}\Lambda _{k})}}\vert coh(\Lambda _{k})\rangle
_{k}=e^{-\frac{|\Lambda _{k}|^{2}}{2k}}\exp (\Lambda _{k}\frac{%
a_{k}^{\dagger }}{k})\vert 0\rangle _{k}.
\end{equation}%
The subscript $k$ denotes that the state is in mode-$k$ subspace $\mathcal{H}%
_{k}$. The $\mathcal{N}(\Lambda _{k})$ is a normalization factor $\langle
0\vert _{k}\exp (\bar{\Lambda}_{k}\frac{a_{k}}{k})\exp (\Lambda _{k}\frac{%
a_{k}^{\dagger }}{k})\vert 0\rangle _{k}=e^{\frac{|\Lambda _{k}|^{2}}{k}}$.
The density matrix of a single-mode pure coherent state is%
\begin{equation}
\rho _{coh(\Lambda _{k})}=\frac{1}{\mathcal{N}(\Lambda _{k})}\exp (\Lambda
_{k}\frac{a_{k}^{\dagger }}{k})\vert 0\rangle _{k}\langle 0\vert _{k}\exp (%
\bar{\Lambda}_{k}\frac{a_{k}}{k})
\end{equation}%
and $\mathrm{tr}~\rho _{coh(\Lambda _{k})}=1$.

The normalization factor of the multi-mode coherent state is as follows. The
normalization factor for $\vert Coh(\{\Lambda _{k}\})\rangle $ is
\begin{equation}
\mathcal{N}(\{\Lambda _{k}\})=\langle Coh(\{\Lambda _{k}\})|Coh(\{\Lambda
_{k}\})\rangle =\exp (\sum\limits_{k=1}^{\infty }\frac{|\Lambda _{k}|^{2}}{k}%
).
\end{equation}

A \emph{special} multi-mode\emph{\ }coherent state is when $\Lambda
_{k}:=(\Lambda )^{k}$, in which $\Lambda \in \mathbb{C}$. The amplitude in
each mode is correlated, since $\Lambda _{k}=\Lambda ^{k}$. This multi-mode
state is
\begin{equation}
\vert Coh(\Lambda )\rangle =\vert Coh(\{\Lambda ^{k}\})\rangle
=\prod_{k=1}^{\infty }\exp (\Lambda ^{k}\frac{a_{k}^{\dagger }}{k})\vert
0\rangle _{k}=\prod_{k=1}^{\infty }\vert coh(\Lambda ^{k})\rangle .
\end{equation}%
We denote $\rho (\Lambda )=\rho (\{\Lambda ^{k}\})$ and the normalization
factor $\mathcal{N}(\Lambda )=\mathcal{N}(\{\Lambda ^{k}\})$. The
normalization for $\vert Coh(\Lambda )\rangle ~$is%
\begin{equation}
\mathcal{N}(\Lambda )=\langle Coh(\Lambda )|Coh(\Lambda )\rangle =\exp
(\sum\limits_{k=1}^{\infty }\frac{|\Lambda |^{2k}}{k})=\frac{1}{(1-|\Lambda
|^{2})}.
\end{equation}%
The expansion we use is $-\mathrm{ln}(1-x)=\sum\limits_{k=1}^{\infty }\frac{%
x^{k}}{k}$. The density matrix for the special multi-mode coherent state is
\begin{equation}
\rho (\Lambda )=\frac{1}{\mathcal{N}(\Lambda )}\prod\limits_{k=1}^{\infty
}\exp (\Lambda ^{k}\frac{a_{k}^{\dagger }}{k})\vert 0\rangle _{k}\langle
0\vert _{k}\exp (\bar{\Lambda}^{k}\frac{a_{k}}{k}).
\label{density matrix Coh}
\end{equation}%
The product is tensor product and $\mathrm{tr}~\rho (\Lambda )=1$.

The inner product of two multi-mode coherent states is
\begin{eqnarray}
\langle Coh(\Lambda _{(0)})|Coh(\Lambda _{(1)})\rangle
&=&\prod\limits_{k=1}^{\infty }\langle 0\vert _{k}\exp (\bar{\Lambda}%
_{(0)}^{k}\frac{a_{k}}{k})\exp (\Lambda _{(1)}^{k}\frac{a_{k}^{\dagger }}{k}%
)\vert 0\rangle _{k}  \notag \\
&=&\frac{1}{(1-\bar{\Lambda}_{(0)}\Lambda _{(1)})}.
\end{eqnarray}%
Hence the inner product of two unit-norm multi-mode coherent states is
\begin{eqnarray}
&&\frac{1}{\sqrt{\mathcal{N}(\Lambda _{(0)})\mathcal{N}(\Lambda _{(1)})}}%
\langle Coh(\Lambda _{(0)})|Coh(\Lambda _{(1)})\rangle  \notag \\
&=&\frac{\left( 1-|\Lambda _{(0)}|^{2}\right) ^{1/2}\left( 1-|\Lambda
_{(1)}|^{2}\right) ^{1/2}}{(1-\bar{\Lambda}_{(0)}\Lambda _{(1)})}.
\end{eqnarray}

\subsection{Phase shift operators}

\label{sec 2.2}

In this section, we consider unitary operations or unitary transformations
on the states considered throughout this paper, in a general setting. We
first discuss the unitary operations on the generalized coherent states.
These unitary operations act similarly on other classes of states that we
shall discuss in Sec. \ref{sec 3} and Sec. \ref{sec 5}.

We first define a generalized phase shift operator. Consider the ordinary
phase shift operator $\exp (i\theta _{k}\hat{N}_{k})=\exp (i\frac{\theta _{k}%
}{k}a_{k}^{\dagger }a_{k})$, which acts on the single-mode coherent state as%
\begin{equation}
\exp (i\frac{\theta _{k}}{k}a_{k}^{\dagger }a_{k})|coh(\Lambda _{k})\rangle
=|coh(\Lambda _{k}e^{i\theta _{k}})\rangle .
\end{equation}%
Note that
\begin{equation}
\exp (i\theta _{k}\hat{N}_{k})|t_{k}^{l}\rangle =\exp (il\theta
_{k})|t_{k}^{l}\rangle =\exp (ikl\theta )|t_{k}^{l}\rangle .
\end{equation}%
Consider%
\begin{equation}
\theta _{k}=k\theta .
\end{equation}%
In this case, for the multi-mode coherent state $|Coh(\Lambda )\rangle $, $%
\Lambda _{k}e^{i\theta _{k}}=\Lambda ^{k}e^{ik\theta }$.$~$The phase shift
operator on the multi-mode coherent state is then%
\begin{equation}
\exp (\sum\limits_{k}i\frac{\theta _{k}}{k}a_{k}^{\dagger
}a_{k})|Coh(\Lambda )\rangle =|Coh(\Lambda e^{i\theta })\rangle .
\end{equation}

Hence we define
\begin{equation}
\exp (i\theta \hat{E}):=\exp (\sum\limits_{k}i\theta a_{k}^{\dagger }a_{k})
\end{equation}%
to be the phase shift operator on $|Coh(\Lambda )\rangle $,$\ $where $\hat{E}%
=\sum_{k}a_{k}^{\dagger }a_{k}$ is the excitation energy operator. The $%
\theta $ can be viewed as a variable conjugate to the generator $\hat{E}$.
The action is%
\begin{equation}
\exp (i\theta \hat{E})|Coh(\Lambda )\rangle =|Coh(\Lambda e^{i\theta
})\rangle
\end{equation}%
for any $|Coh(\Lambda )\rangle $. For any superposition of coherent states,
the action is
\begin{equation}
\exp (i\theta \hat{E})(c_{1}|Coh(\Lambda _{1})\rangle +c_{2}|Coh(\Lambda
_{2})\rangle )=c_{1}|Coh(\Lambda _{1}e^{i\theta })\rangle +c_{2}|Coh(\Lambda
_{2}e^{i\theta })\rangle .
\end{equation}%
An interesting superposition is
\begin{equation}
|Coh(\Lambda )\rangle +e^{i\varphi }|Coh(\Lambda e^{i\chi })\rangle
=(1+e^{i\varphi I}e^{i\chi \hat{E}})|Coh(\Lambda )\rangle .
\end{equation}

The phase shift operator can be used to define a generating function for$~$a
\emph{general} state $|\Psi \rangle ,$%
\begin{equation}
\langle \exp (i\theta \hat{E})\rangle _{|\Psi \rangle }:=\frac{\langle \Psi
|\exp (i\theta \hat{E})|\Psi \rangle }{\Vert \Psi \Vert ^{2}}.
\end{equation}%
We define the expectation value of the excitation energy
\begin{equation}
E_{|\Psi \rangle }:=\langle \hat{E}\rangle _{|\Psi \rangle }
\end{equation}%
above the ground state, and the zero point energy is subtracted out in the
definition of $\hat{E}$. In other words, for ground state, $E$ is the energy
subtracting the zero point energy. For the special case $|Coh(\Lambda
)\rangle $, its generating function is%
\begin{equation}
\langle \exp (i\theta \hat{E})\rangle _{Coh(\Lambda )}=\frac{(1-|\Lambda
|^{2})}{(1-|\Lambda |^{2}e^{i\theta })}.
\end{equation}%
By taking derivatives, we have $\langle \hat{E}\rangle _{Coh(\Lambda )}=%
\frac{|\Lambda |^{2}}{1-|\Lambda |^{2}}$, which is the same as the direct
summation in Sec. \ref{sec 4.1} as in (\ref{exitation energy}).

We define the variance of the excitation energy $\hat{E}$ on a general
state,~to be $(\Delta E)^{2}$, by%
\begin{equation}
(\Delta E)_{|\Psi \rangle }^{2}:=\langle \hat{E}^{2}\rangle _{|\Psi \rangle
}-\langle \hat{E}\rangle _{|\Psi \rangle }^{2}.
\end{equation}

For a general state $|\Psi \rangle $, $E_{|\Psi \rangle }=\langle \hat{E}%
\rangle _{|\Psi \rangle }$ and $(\Delta E)_{|\Psi \rangle }^{2}$ are two
important physical quantities charactering that state. These two quantities
can be calculated by the generating function $\langle \exp (i\theta \hat{E}%
)\rangle _{|\Psi \rangle }$. Note that $\theta $ is also the phase variable
or the angular variable of the phase space in our case.

Using Heisenberg uncertainty relation, $\theta $ is the conjugate variable
for the generator $\hat{E}$ in $\exp (i\theta \hat{E})$, and hence we are
subject to a Heisenberg uncertainty relation between $\Delta E$ and $\Delta
\theta $, i.e. $\Delta E\cdot \Delta \theta \gtrsim $ $\frac{1}{2}$.

The $(\Delta E)^{2}$ is related to quantum Fisher information (QFI) of the
state. The quantum Fisher information $4(\Delta E)^{2}$ characterizes how
well the state can be localized in the $\theta $ direction, or how well the
state can be resolved and distinguished in the $\theta $ direction. The
concept of quantum Fisher information is useful for precision quantum
measurement \cite{Braunstein,Hyllus,Toth}. In quantum information theory,
the quantum Fisher information of the state $|\Psi \rangle $ can be written
as
\begin{equation}
\mathrm{QFI}(|\Psi \rangle )=4(\Delta E)_{|\Psi \rangle }^{2}.
\end{equation}%
And $\Delta \theta _{|\Psi \rangle }\gtrsim $ $\frac{1}{2\Delta E_{|\Psi
\rangle }}=\frac{1}{\sqrt{\mathrm{QFI}(|\Psi \rangle )}}\mathrm{.}$ Small $%
\Delta \theta _{|\Psi \rangle }$ means that the state is localized on the $%
\theta $ direction, such as $|Coh(\Lambda )\rangle $, and big $\Delta \theta
_{|\Psi \rangle }$ means that the state is delocalized on the $\theta $
direction, such as $|t_{k}^{l}\rangle $. States with symmetry on $\theta $
direction has big $\Delta \theta _{|\Psi \rangle }$. A state with a greater $%
4(\Delta E)^{2}$ has a greater localizabiliy and resolution in the $\theta ~$%
direction. Hence, $4(\Delta E)_{|\Psi \rangle }^{2}$ measures the quantum
Fisher information of the state and the localizability of the state in the
angular direction in phase space. We shall calculate $(\Delta E)_{Cat_{\pm
}(\Lambda )}^{2}$,$\mathrm{~}(\Delta E)_{~\Phi _{N,n}(\Lambda )}^{2}$,$%
~(\Delta E)_{cat_{\pm }(\Lambda _{k})}^{2}$ of the states in Sec. \ref{sec 3}
and Sec. \ref{sec 5}, also because this quantity shows the energy
fluctuation of the underlying superposition state. We can, in particular,
focus on states with large quantum Fisher information.

\subsection{Parity operators}

We now define a parity operator associated to this phase shift operator in
Sec. \ref{sec 2.2}. The parity operator is a special case of the phase shift
operator when the phase rotation is $\pi $. The parity operator is thus%
\begin{equation}
\hat{P}=\exp (\sum\limits_{k}i\pi a_{k}^{\dagger }a_{k})=\exp (i\pi \hat{E}),
\end{equation}%
and
\begin{equation}
\hat{P}|Coh(\Lambda )\rangle =|Coh(-\Lambda )\rangle ,
\end{equation}%
for any coherent state and their superpositions thereof. We also have $\hat{P%
}^{2}=I$ and the following relations
\begin{equation}
\hat{P}=\prod_{k=1}^{\infty }(\hat{P}_{k})^{k},\ \ \ \ \ \ \text{\textrm{\ \
\ }}\ \hat{P}_{k}=\exp (i\frac{\pi }{k}a_{k}^{\dagger }a_{k}),
\end{equation}%
\begin{equation}
\ \hat{P}_{k}|coh(\Lambda _{k})\rangle =|coh(-\Lambda _{k})\rangle ,~\ ~~\ \
~~(\hat{P}_{k})^{2}=I.~\ \ \ \
\end{equation}%
For$~$a general state $|\Psi \rangle ,$%
\begin{equation}
\langle \hat{P}\rangle _{|\Psi \rangle }:=\frac{\langle \Psi |\hat{P}|\Psi
\rangle }{\Vert \Psi \Vert ^{2}}.
\end{equation}%
For the coherent state,
\begin{equation}
\langle \hat{P}\rangle _{Coh(\Lambda )}=\frac{1-|\Lambda |^{2}}{1+|\Lambda
|^{2}}.
\end{equation}%
Note $0\leq \langle \hat{P}\rangle _{Coh(\Lambda )}\leq 1$ which is an
interesting property, and%
\begin{equation}
(\Delta P)_{Coh(\Lambda )}^{2}=\frac{4|\Lambda |^{2}}{(1+|\Lambda |^{2})^{2}}%
.
\end{equation}%
For a general state $|\Psi \rangle $, because$~\hat{P}$ is a Hermitian
operator, it is easy to infer that
\begin{equation}
-1\leq \langle \hat{P}\rangle _{_{|\Psi \rangle }}\leq 1.
\end{equation}

\subsection{Multi-parameter multi-mode coherent states and the merging of
two bumps}

The multi-parameter coherent states can be defined as
\begin{equation}
\vert \Psi \rangle =\vert Coh(x_{1},x_{2},\dots ,x_{m})\rangle
=B_{+}(x_{1},x_{2},...,x_{m})\lvert {0}\rangle
:=\prod_{i=1}^{m}B_{+,x_{i}}\lvert {0}\rangle ,
\end{equation}%
where $B_{+,x_{i}}=\exp (\sum_{k}x_{i}^{~k}\frac{a_{k}^{\dagger }}{k})$, and
$\vert Coh(x_{1},x_{2},\dots ,x_{m})\rangle $ corresponds to $m$ bumps \cite%
{Berenstein:2017abm,Lin:2017dnz} near the edge of the black disk. The shape
of the bumps is calculated by the expectation value \cite{Berenstein:2017abm}
of the chiral field $\langle \hat{\phi}(\theta )\rangle _{\vert \Psi \rangle
}$, where $\hat{\phi}(\theta )=\sum_{k>0}(a_{k}\exp (-ik\theta
)+a_{k}^{\dagger }\exp (ik\theta ))$. Each bump is localized in the angular
direction $\mathrm{\arg }(x_{i})$.

Consider that the two bumps with an angular distance seperation $2\theta $,
merges into a single bump. Consider states $\vert \Psi _{1}\rangle =\vert
Coh(\Lambda _{1},\Lambda _{2})\rangle $ and $\vert \Psi _{2}\rangle =\vert
Coh(\Lambda _{3})\rangle $, where $\Lambda _{1}=\Lambda $, $\Lambda
_{2}=\Lambda e^{2i\theta }$, $\Lambda _{3}=\frac{2|\Lambda |}{1+|\Lambda
|^{2}}\Lambda e^{i\theta }$. So that the total excitation energy of the two
states is the same, i.e. $E=\langle \hat{E}\rangle _{\vert \Psi _{1}\rangle
}=\langle \hat{E}\rangle _{\vert \Psi _{2}\rangle }$, where $\hat{E}%
=\sum_{k}k\hat{N}_{k}$. We assume that the angle $\theta $ is very small for
simplicity of the calculation. In the context of gauge/gravity duality,
these two states belong to the same Hilbert space and have the same
excitation energy, measured at the asymptotic infinity of the spacetime. The
transition amplitude calculated by the field theory side between these two
states is%
\begin{equation}
A=\frac{\langle \Psi _{1}|\Psi _{2}\rangle }{\vert \Psi _{1}\vert \vert \Psi
_{2}\vert }\simeq \frac{1}{E^{\frac{1}{2}}}\frac{16(1+E^{2}\sin ^{2}\theta
)^{\frac{1}{2}}}{9(1+\frac{16}{9}E^{2}\sin ^{2}\frac{\theta }{2})},
\end{equation}%
for large $E\gg 1$. This expression has taken into account small $\theta $
and small $E^{-1}$ limits. This is the transition from a pair of bumps to a
single bump that the two merged into. The above amplitude is a real number,
because we have made the final state to be in the middle of the angular
separation, for simplicity of the calculation.

By the map of gauge/gravity duality, this is a transition amplitude between
two gravitational states. Transition amplitudes between backreacted
geometries were also calculated in \cite{Berenstein:2017abm,Brown:2006zk,Diaz:2015tda,Lin:2017dnz}. Similar computations
of tunneling probabilities between different fuzzball microstate geometries
were also computed in the context of quantum gravity in \cite{Mathur:2008kg}.

\section{Superposition states and cat states}

\label{sec 3} \renewcommand{\theequation}{3.\arabic{equation}} %
\setcounter{equation}{0} \renewcommand{\thethm}{3.\arabic{thm}} %
\setcounter{thm}{0} \renewcommand{\theprop}{3.\arabic{prop}} %
\setcounter{prop}{0}

\subsection{Superposition states and Schrodinger cat states}

\label{sec 3.1}

These coherent states in Sec. \ref{sec 2.1} are in the same Hilbert space,
hence we can perform superpositions of these states. We can also perform
various other quantum operations \cite{Quantum information} on these states.
Consider the Schrodinger cat state%
\begin{eqnarray}
|Cat_{\pm }(\Lambda )\rangle &=&\frac{1}{\sqrt{N_{\pm }}}(|Coh(\Lambda
)\rangle \pm |Coh(-\Lambda )\rangle ) \\
&=&\frac{1}{\sqrt{N_{\pm }}}(1\pm \hat{P})|Coh(\Lambda )\rangle ,
\end{eqnarray}%
where $|Cat_{\pm }(\Lambda )\rangle $ has unit norm, and
\begin{equation}
N_{+}=4(1-|\Lambda |^{4})^{-1}~,~N_{-}=4|\Lambda |^{2}(1-|\Lambda
|^{4})^{-1}.
\end{equation}%
The $|Cat_{\pm }(\Lambda )\rangle $ are orthogonal to each other, i.e. $%
\langle Cat_{-}(\Lambda )|Cat_{+}(\Lambda )\rangle =0$. We have that $%
\langle \hat{P}\rangle _{|Cat_{\pm }(\Lambda )\rangle }=\pm 1$.$~$They form
orthogonal basis of a qubit, and these states can be used to construct
qubits. In the context of gauge/gravity duality, on the gravity side, the
cat states mean that there is a half percent change that there is a bump at
one end of a black disk, and another half percent chance that there is a
bump at the opposite end of the black disk.

The Schrodinger cat states are the superposition states of macroscopic
quantum states, which was proposed in thought experiment by Schrodinger \cite%
{Schrodinger:1935zz}. These cat states are similar to the superposition of
photonic coherent states, in e.g. \cite{Andersen etal,Nielsen etal}.
Such Schrodinger cat states have been prepared in photonic experiments, see
\cite{Andersen etal,Nielsen etal} and references therein.

The generating function of the cat states is
\begin{equation}
\langle \exp (i\theta \hat{E})\rangle _{Cat_{\pm }(\Lambda )}=\frac{1}{%
N_{\pm }}(\frac{2}{1-e^{i\theta }|\Lambda |^{2}}\pm \frac{2}{1+e^{i\theta
}|\Lambda |^{2}}).
\end{equation}%
The excitation energy is given by
\begin{eqnarray}
\langle \hat{E}\rangle _{Cat_{+}(\Lambda )} &=&\frac{2|\Lambda |^{4}}{%
(1-|\Lambda |^{2})(1+|\Lambda |^{2})}, \\
\langle \hat{E}\rangle _{Cat_{-}(\Lambda )} &=&\frac{1+|\Lambda |^{4}}{%
(1-|\Lambda |^{2})(1+|\Lambda |^{2})},
\end{eqnarray}%
and
\begin{equation}
(\Delta E)_{Cat_{\pm }(\Lambda )}^{2}=\frac{4|\Lambda |^{4}}{(1-|\Lambda
|^{2})^{2}(1+|\Lambda |^{2})^{2}}.
\end{equation}

Now we consider entanglement between \emph{light} states and \emph{heavy}
states. In the context of gauge/gravity duality, the heavy states can be
viewed as backgrounds and light states can be viewed as probes. Consider the
EPR state \cite{Einstein:1935rr}%
\begin{equation}
\Psi =\frac{1}{\sqrt{2}}(|\phi _{+}\rangle |Cat_{+}(\Lambda )\rangle +|\phi
_{-}\rangle |Cat_{-}(\Lambda )\rangle ),
\end{equation}%
where $|\phi _{+}\rangle ,|\phi _{-}\rangle $ have unit norm and are
orthogonal to each other, i.e. $\langle \phi _{-}|\phi _{+}\rangle =0$.
Here, $|\phi _{+}\rangle ,|\phi _{-}\rangle $ are light states and $%
|Cat_{+}(\Lambda )\rangle $,$~|Cat_{-}(\Lambda )\rangle $ are heavy states.

For example, we can make graviton number states, involving only one mode,
\begin{equation}
|\phi _{+}\rangle =\frac{1}{\sqrt{k^{l_{1}}l_{1}!}}|{t_{k}^{l_{1}}}\rangle
,\ \ \ |\phi _{-}\rangle =\frac{1}{\sqrt{k^{l_{2}}l_{2}!}}|{t_{k}^{l_{2}}}%
\rangle ,  \label{state 01}
\end{equation}%
with $l_{1}\neq l_{2}$.

In the above case, each term, such as $\frac{1}{\sqrt{k^{l_{i}}l_{i}!}}|{%
t_{k}^{l_{i}}}\rangle |Coh(\Lambda )\rangle $, can be viewed as an
excitation of a light state, e.g. $|{t_{k}^{l_{i}}}\rangle $, on the
background of a heavy state, e.g. $|Coh(\Lambda )\rangle $. In this term,
the background is a bump at the edge of the disk, and the light state is $%
l_{i}$ Kaluza-Klein (KK) gravitons each with momentum $k$. This state is
similar to the micro-macro entangled states for photonic coherent states as
in e.g. \cite{Andersen etal}.

Another example, involving two modes, is%
\begin{equation}
|\phi _{+}\rangle =\frac{1}{\sqrt{k^{l_{1}}l_{1}!}}|{t_{k_{1}}^{l_{1}}}%
\rangle |{t_{k_{2}}^{0}}\rangle ,\ \ \ |\phi _{-}\rangle =\frac{1}{\sqrt{%
k^{l_{2}}l_{2}!}}|{t_{k_{1}}^{0}}\rangle |{t_{k_{2}}^{l_{2}}}\rangle ,
\end{equation}%
with $k_{1}\neq k_{2}$, $l_{1}\neq 0$, $l_{2}\neq 0$. A special case,
involving two modes, is $\ $%
\begin{equation}
|\phi _{+}\rangle =\frac{1}{\sqrt{k_{1}}}|{t_{k_{1}}^{1}}\rangle |{%
t_{k_{2}}^{0}}\rangle ,\ \ \ |\phi _{-}\rangle =\frac{1}{\sqrt{k_{2}}}|{%
t_{k_{1}}^{0}}\rangle |{t_{k_{2}}^{1}}\rangle ,
\end{equation}%
with $k_{1}\neq k_{2}.$

\subsection{$N$-state Schrodinger cat}

Now we consider the superposition of $N$ distinct macroscopic quantum
states, generalizing the superposition of only two states in the preceding
section \ref{sec 3.1}. Define the $N$-state Schrodinger cat
\begin{equation}
\Phi _{N,n}(\Lambda )=\frac{1}{\sqrt{\mathcal{N}_{N,n}}}%
\sum_{m=0}^{N-1}|Coh(\Lambda e^{i\frac{2\pi }{N}m})\rangle e^{-i\frac{2\pi }{%
N}mn}.  \label{N state cat}
\end{equation}%
for $n=0,...,N-1$. In this superposition, there are $N$ multi-mode coherent
states. We make $\Vert \Phi _{N,n}(\Lambda )\Vert =1$ and require $%
0<|\Lambda |<1$.$~$The normalization factor is%
\begin{eqnarray}
\mathcal{N}_{N,n} &=&\sum_{0\leq m,m^{\prime }\leq N-1}(1-|\Lambda |^{2}e^{i%
\frac{2\pi }{N}(m-m^{\prime })})^{-1}e^{-i\frac{2\pi }{N}(m-m^{\prime })n}
\notag \\
&=&\sum_{k=0}^{\infty }(|\Lambda |^{2k}\sum_{0\leq m,m^{\prime }\leq N-1}e^{i%
\frac{2\pi }{N}(k-n)(m-m^{\prime })})=\frac{N^{2}|\Lambda |^{2n}}{1-|\Lambda
|^{2N}},  \label{normalization factor}
\end{eqnarray}%
where in the last line we used that
\begin{equation}
\sum_{m,m^{\prime }=0}^{N-1}e^{i\frac{2\pi }{N}(m-m^{\prime })(k-n)}=\left\{
\begin{array}{cc}
N^{2} & \;k-n\equiv 0~\func{mod}~N \\
0 & \mathrm{otherwise}%
\end{array}%
\right. .
\end{equation}

There is a discrete symmetry of this state, which is the symmetry of the
finite group $Z_{N}$. The state $|\Phi _{N,n}(\Lambda )\rangle $ is the $n$%
-th irreducible representation of the finite group $Z_{N}$. The action of $%
Z_{N}$ is generated by $\tau :\theta \rightarrow \theta +\frac{2\pi }{N}$.

The action of the phase shift operator is:
\begin{equation}
\exp (i\theta \hat{E})~\Phi _{N,n}(\Lambda )=\Phi _{N,n}(\Lambda e^{i\theta
}).
\end{equation}%
The $Z_{N}$ action $\tau :\theta \rightarrow \theta +\frac{2\pi }{N}$ is:%
\begin{equation}
\tau ~\Phi _{N,n}(\Lambda )=e^{-i\frac{2\pi }{N}n}\Phi _{N,n}(\Lambda e^{i%
\frac{2\pi }{N}}).~~~~~~
\end{equation}%
Hence%
\begin{equation}
\tau =e^{-i\frac{2\pi }{N}nI}e^{i\frac{2\pi }{N}\hat{E}}.
\end{equation}%
The action of the parity operator is:
\begin{equation}
\hat{P}~\Phi _{N,n}(\Lambda )=(-1)^{n}\Phi _{N,n}(\Lambda )=\Phi
_{N,n}(-\Lambda ).
\end{equation}%
Hence for even $N$, we have $(-1)^{n}\tau ^{N/2}~\Phi _{N,n}(\Lambda )=\hat{P%
}~\Phi _{N,n}(\Lambda )$, and%
\begin{equation}
(-1)^{n}\tau ^{N/2}=\hat{P}.
\end{equation}%
There is a periodicity in $n$, for $n^{\prime }=n$ mod $N$, they are the
equivalent state. Hence $n=0,...,N-1$.$~$Note that this $N$ does not denote
the rank of the gauge group.

This state includes the states in preceding sections \ref{sec 2.2} and \ref%
{sec 3.1} as the special cases for small $N$. For $N=1$,
\begin{equation}
\Phi _{1,n}(\Lambda )=\frac{1}{\sqrt{\mathcal{N}_{1,n}}}|Coh(\Lambda
)\rangle ,
\end{equation}%
and $n=0$. For $N=2$,
\begin{equation}
\Phi _{2,n}(\Lambda )=\frac{1}{\sqrt{\mathcal{N}_{2,n}}}(|Coh(\Lambda
)\rangle +(-1)^{n}|Coh(-\Lambda )\rangle ).
\end{equation}%
and $n=0,1$. So $\Phi _{2,n=0}=|Cat_{+}(\Lambda )\rangle $, $\Phi
_{2,n=1}=|Cat_{-}(\Lambda )\rangle $.

In the large $N$ limit, $\theta _{m}=\frac{2\pi }{N}m$,$~\lim_{N\rightarrow
\infty }\frac{1}{N}\sum_{m=0}^{N-1}=\frac{1}{2\pi }\int_{0}^{2\pi }d\theta $%
. Hence
\begin{equation}
\lim_{N\rightarrow \infty }\Phi _{N,n}(\Lambda )=\frac{1}{\sqrt{\mathcal{N}%
_{n}}}\frac{N}{2\pi }\int_{0}^{2\pi }d\theta e^{-in\theta }|Coh(\Lambda
e^{i\theta })\rangle .
\end{equation}%
Here $\mathcal{N}_{n}=\lim_{N\rightarrow \infty }\mathcal{N}%
_{N,n}=N^{2}|\Lambda |^{2n}$.

For general $N,n$,
\begin{equation}
\langle \exp (i\theta \hat{E})\rangle _{\Phi _{N,n}(\Lambda )}=\frac{%
e^{in\theta }(1-|\Lambda |^{2N})}{1-|\Lambda |^{2N}e^{iN\theta }},
\end{equation}%
\begin{equation}
\langle \hat{E}\rangle _{\Phi _{N,n}(\Lambda )}=\frac{n+(N-n)|\Lambda |^{2N}%
}{1-|\Lambda |^{2N}},
\end{equation}%
\begin{equation}
(\Delta E)_{~\Phi _{N,n}(\Lambda )}^{2}=\frac{N^{2}|\Lambda |^{2N}}{%
(1-|\Lambda |^{2N})^{2}}.  \label{delta E^2 02}
\end{equation}

We denote the Young tableau state with a single row of length $n$ as $\lvert
{\triangle _{n}}\rangle ~$\cite{Berenstein:2017abm,Lin:2017dnz}. This
state can be written as Schur polynomial operators \cite{Corley:2001zk}. We
have the following proposition:

\begin{prop}
The generalized Schrodinger cat state $|\Phi _{N,n}(\Lambda )\rangle $
approaches the Young tableau state $\lvert {\triangle _{n}}\rangle $ with
fidelity approaching $1$, in the large $N$ limit.
\end{prop}

{\textbf{\textit{Proof}}}. In the large $N$ limit, $\theta _{m}=\frac{2\pi }{%
N}m$,$~\lim_{N\rightarrow \infty }\frac{1}{N}\sum_{m=0}^{N-1}=\frac{1}{2\pi }%
\int_{0}^{2\pi }d\theta $. Hence
\begin{equation}
\lim_{N\rightarrow \infty }\Phi _{N,n}(\Lambda )=\frac{1}{\sqrt{\mathcal{N}%
_{n}}}\frac{N}{2\pi }\int_{0}^{2\pi }d\theta e^{-in\theta }|Coh(\Lambda
e^{i\theta })\rangle ,  \label{N-cat}
\end{equation}%
where $\mathcal{N}_{n}=\lim_{N\rightarrow \infty }\mathcal{N}%
_{N,n}=N^{2}|\Lambda |^{2n}$.$~$We have $\Vert \Phi _{N,n}(\Lambda )\Vert =1$%
. The one-row Young tableau state is \cite{Lin:2017dnz}%
\begin{equation}
\lvert {\triangle _{n}}\rangle =\frac{1}{2\pi i}\oint_{\mathcal{C}}\frac{%
\mathrm{d}w}{w}w^{-n}|Coh(w)\rangle ,
\end{equation}%
where $\mathcal{C}$ can be any path that encloses $0$. The$\lvert {\triangle
_{n}}\rangle $ has unit norm. It has property
\begin{equation}
|Coh(w)\rangle =\sum_{n}w^{n}\lvert {\triangle _{n}}\rangle ,~~~\langle
\Delta _{n}|Coh(w)\rangle =w^{n}.  \label{inner product}
\end{equation}%
We define $w=\Lambda e^{i\theta }=|\Lambda |e^{i\theta +i\theta _{0}}$.
Hence
\begin{equation}
\lim_{N\rightarrow \infty }\Phi _{N,n}(\Lambda )=\frac{1}{\sqrt{\mathcal{N}%
_{n}}}N|\Lambda |^{n}e^{in\theta _{0}}|\Delta _{n}\rangle .
\end{equation}%
Alternatively, we compute the inner product between $|\Phi _{N,n}(\Lambda
)\rangle $ and $|\Delta _{n}\rangle $, and then take the large $N$ limit. By
(\ref{inner product}),%
\begin{equation}
\langle \Delta _{n}|\Phi _{N,n}(\Lambda )\rangle =\frac{1}{\sqrt{\mathcal{N}%
_{N,n}}}\sum_{m=0}^{N-1}\Lambda ^{n}=\frac{N\Lambda ^{n}}{\sqrt{\mathcal{N}%
_{N,n}}}.
\end{equation}%
The fidelity between $|\Phi _{N,n}(\Lambda )\rangle $ and $|\Delta
_{n}\rangle $ is
\begin{equation}
F=|\langle \Delta _{n}|\Phi _{N,n}(\Lambda )\rangle |=\frac{N|\Lambda |^{n}}{%
\sqrt{\mathcal{N}_{N,n}}}.
\end{equation}%
Hence, we find that $\lim_{N\rightarrow \infty }\Phi _{N,n}(\Lambda
)=e^{in\theta _{0}}|\Delta _{n}\rangle $, $\lim_{N\rightarrow \infty
}\langle \Delta _{n}|\Phi _{N,n}(\Lambda )\rangle =e^{in\theta _{0}}$ and
the fidelity $F~$approaches $1$. Hence, at large $N$, the $N$-state
schrodinger cat state approaches the one-row Young tableau state, with
fidelity between the two states asymptotically reaches $1$.{\hfill $\square $%
}\newline

The fact that the $N$-state cat has limit to the Young tableau state, is of
significance. The superposition of the angularly localized states produces
an angularly de-localized state. From the behavior of Eq. (\ref{delta E^2 02}%
) with respect to $N$, we see that the state $\Phi _{N,n}(\Lambda )$ with a
bigger $N$ is more de-localized than the state with a smaller $N$. In the
large $N$, it approaches a state very different from each typical state in
the superposition. This is another very interesting example of a
superposition pure state. Other aspects of the superposition states were
discussed in \cite{Berenstein:2017abm,Lin:2017dnz}.

\section{Noisy coherent states}

\label{sec 4} \renewcommand{\theequation}{4.\arabic{equation}} %
\setcounter{equation}{0} \renewcommand{\thethm}{4.\arabic{thm}} %
\setcounter{thm}{0} \renewcommand{\theprop}{4.\arabic{prop}} %
\setcounter{prop}{0}

\subsection{A class of single mode and multi mode noisy coherent states}

\label{sec 4.1}

Now we consider noisy coherent states$~$with our generalization. These are
generalizations of the original construction of the noisy coherent states of
photons \cite{Helstrom a}. The noisy coherent states were originally
constructed directly using the $P$-representation of the density matrix
operator \cite{Helstrom a}. In quantum information theory, the noisy
coherent states can be generated experimently from pure coherent states
through the action of Gaussian additive noise. We use the creation and
annihilation operators in our case to derive the state to illustrate its
relation to the specific class of coherent states using multi traces in Sec. %
\ref{sec 2.1}.

Now we derive this mixed density matrix, by adding random multi trace
states. Consider $|coh(\Lambda _{k})\rangle _{k}$ which is a pure coherent
state for mode $k$. We define $a_{k}=d_{k}+\Lambda _{k}$, $a_{k}^{\dagger
}=d_{k}^{\dagger }+\bar{\Lambda}_{k}$, and $\frac{1}{k}[d_{k},~d_{k}^{%
\dagger }]=1$. Hence,%
\begin{eqnarray}
d_{k}|coh(\Lambda _{k})\rangle _{k} &=&0, \\
d_{k}|coh(\Gamma _{k})\rangle _{k} &=&(\Gamma _{k}-\Lambda _{k})|coh(\Gamma
_{k})\rangle _{k},
\end{eqnarray}%
and $\frac{1}{\mathcal{N}(\Lambda _{k})}\langle coh(\Gamma
_{k})|_{k}(d_{k}^{\dagger })^{l}(d_{k})^{l}|coh(\Gamma _{k})\rangle
_{k}=|\Gamma _{k}-\Lambda _{k}|^{2l}$. We define%
\begin{equation}
~|t_{k}^{l}(\Lambda _{k})\rangle =\frac{1}{\sqrt{\mathcal{N}(\Lambda _{k})}}%
(d_{k}^{\dagger })^{l}|coh(\Lambda _{k})\rangle _{k}.
\end{equation}%
The state $|t_{k}^{l}(\Lambda _{k})\rangle $ can be viewed as the product of
the multi trace with coherent state. The conjugate is $\langle \bar{t}%
_{k}^{l}(\bar{\Lambda}_{k})|=\frac{1}{\sqrt{\mathcal{N}(\Lambda _{k})}}%
\langle coh(\Lambda _{k})|_{k}(d_{k})^{l}$. Its special cases are $%
|t_{k}^{l}(0)\rangle =|t_{k}^{l}\rangle $ and $|t_{k}^{0}(\Lambda
_{k})\rangle =\frac{1}{\sqrt{\mathcal{N}(\Lambda _{k})}}|coh(\Lambda
_{k})\rangle _{k}$.~The $\hat{n}_{k}=\frac{1}{k}d_{k}^{\dagger }d_{k}$ is
the particle number operator for the $d_{k}^{\dagger }$ excitations, which
can be considered as random thermal particles, and $\hat{n}%
_{k}|t_{k}^{l}(\Lambda _{k})\rangle =l|t_{k}^{l}(\Lambda _{k})\rangle $.
Hence $\langle \hat{n}_{k}\rangle $ is the quantity that measures the
average thermal particle number.

Consider a mixed state density matrix,
\begin{eqnarray}
\rho (\Lambda _{k})_{\mathrm{mix}} &=&\sum_{l=0}^{\infty }\frac{\langle \hat{%
n}_{k}\rangle ^{l}}{(1+\langle \hat{n}_{k}\rangle )^{l+1}}\frac{%
|t_{k}^{l}(\Lambda _{k})\rangle }{\sqrt{k^{l}l!}}\frac{\langle \bar{t}%
_{k}^{l}(\bar{\Lambda}_{k})|}{\sqrt{k^{l}l!}}  \label{Bose Einstein} \\
&=&\frac{1}{\langle \hat{n}_{k}\rangle }\sum_{l=0}^{\infty }\frac{(-1)^{l}}{%
l!}\frac{(d_{k})^{l}|t_{k}^{0}(\Lambda _{k})\rangle \langle \bar{t}_{k}^{0}(%
\bar{\Lambda}_{k})|(d_{k}^{\dagger })^{l}}{l!k^{l}\langle \hat{n}_{k}\rangle
^{l}} \\
&=&\int d^{2}\Gamma _{k}~\frac{1}{\pi k\langle \hat{n}_{k}\rangle }\exp (-%
\frac{|\Gamma _{k}-\Lambda _{k}|^{2}}{k\langle \hat{n}_{k}\rangle }%
)|t_{k}^{0}(\Gamma _{k})\rangle \langle \bar{t}_{k}^{0}(\Gamma _{k})|.
\end{eqnarray}%
The distribution $p_{k}(l)=\frac{\langle \hat{n}_{k}\rangle ^{l}}{(1+\langle
\hat{n}_{k}\rangle )^{l+1}}$ in (\ref{Bose Einstein}) is a Bose-Einstein
distribution of random multi traces on top of a coherent state. The random
traces are the noise. Hence,
\begin{equation}
\rho (\Lambda _{k})_{\mathrm{mix}}=\int d^{2}\Gamma _{k}\frac{1}{\mathcal{N}%
(\Gamma _{k})}~\frac{1}{\pi \mu _{k}}\exp (-|\Gamma _{k}-\Lambda
_{k}|^{2}/\mu _{k})|coh(\Gamma _{k})\rangle \langle coh(\bar{\Gamma}_{k})|,
\end{equation}%
where%
\begin{equation}
\mu _{k}=k\langle \hat{n}_{k}\rangle .
\end{equation}%
Denote $p(\Gamma _{k},\Lambda _{k})=\frac{1}{\pi \mu _{k}}\exp (-|\Gamma
_{k}-\Lambda _{k}|^{2}/\mu _{k})$, and we have $\int p(\Gamma _{k},\Lambda
_{k})d^{2}\Gamma _{k}=1$. The density operator $\rho (\Lambda _{k})_{\mathrm{%
mix}}$ is a superposition of the density operators with a distribution
function $p(\Gamma _{k},\Lambda _{k})$ in the coherent state basis. This is
a superposition mixed state, instead of a superposition pure state.

The $\hat{N}_{k}=\frac{1}{k}a_{k}^{\dagger }a_{k}$ is the total particle
number operator in mode $k$. The mean particle number in the pure state in
mode $k$ is
\begin{equation}
\langle \hat{N}_{k}\rangle _{coh(\Gamma _{k})}=\frac{1}{\mathcal{N}(\Gamma
_{k})}\langle 0|_{k}\exp (\bar{\Gamma}_{k}\frac{a_{k}}{k})\left( \frac{1}{k}%
a_{k}^{\dagger }a_{k}\right) \exp (\Gamma _{k}\frac{a_{k}^{\dagger }}{k}%
)|0\rangle _{k}=\frac{|\Gamma _{k}|^{2}}{k}.
\end{equation}%
In the mixed state,%
\begin{eqnarray}
\langle \hat{N}_{k}\rangle _{\rho (\Lambda _{k})_{\mathrm{mix}}} &=&\mathrm{%
tr~}(\rho (\Lambda _{k})_{\mathrm{mix}}\frac{1}{k}a_{k}^{\dagger
}a_{k})=\int d^{2}\Gamma _{k}\frac{1}{\pi \mu _{k}}\exp (-|\Gamma
_{k}-\Lambda _{k}|^{2}/\mu _{k})\frac{|\Gamma _{k}|^{2}}{k}  \notag \\
&=&\frac{|\Lambda _{k}|^{2}}{k}+\frac{\mu _{k}}{k}=\langle \hat{N}%
_{k}\rangle _{coh(\Lambda _{k})}+\langle \hat{n}_{k}\rangle .
\end{eqnarray}

The $\hat{E}_{k}=a_{k}^{\dagger }a_{k}$ is the excitation energy operator in
the mode $k$ . The excitation energy in mode $k$ is $E_{k}$.\ The total
excitation energy operator is $\hat{E}=\sum_{k}\hat{E}_{k}=\sum_{k}a_{k}^{%
\dagger }a_{k}$. The generating function of the mixed state is $\langle \exp
(i\theta \hat{E})\rangle _{\rho _{\mathrm{mix}}}=~$\textrm{tr}$(e^{i\theta
\hat{E}}\rho _{\mathrm{mix}})$. We have in mode $k$,%
\begin{equation}
\langle \hat{E}_{k}\rangle =k\langle \hat{N}_{k}\rangle =|\Lambda
_{k}|^{2}+\mu _{k}.
\end{equation}

The fidelity between the pure coherent state $\rho _{\mathrm{pure}}=\rho
_{_{coh(\Lambda _{k})}}~$and noisy coherent state $\rho _{\mathrm{mix}}=\rho
(\Lambda _{k})_{\mathrm{mix}}$ is%
\begin{equation}
\text{\textrm{tr}}(\rho _{\mathrm{mix}}\rho _{\mathrm{pure}})=\frac{1}{%
\mathcal{N}(\Lambda _{k})}\langle coh(\Lambda _{k})\vert \rho (\Lambda
_{k})_{\mathrm{mix}}\vert coh(\Lambda _{k})\rangle =\frac{1}{1+\frac{\mu _{k}%
}{k}}=\frac{1}{1+\langle \hat{n}_{k}\rangle }.
\end{equation}

Consider in each mode $k$, the signal is centered at $\Lambda _{k}$. We
define a special case of multi mode noisy coherent states, by centering the
signal at $\Lambda _{k}$ for each mode $k$. We can introduce the noise in
each mode $k$, in the multi-mode case. This is to introduce distribution in
each mode, resulting in a joint distribution function $P(\{\Gamma _{k}\})~$%
in the joint parameter space,
\begin{equation}
P(\{\Gamma _{k}\})=\prod_{k=1}^{\infty }\frac{1}{\pi \mu _{k}}\exp
(\sum\limits_{k=1}^{\infty }-|\Gamma _{k}-\Lambda _{k}|^{2}/\mu _{k}).
\end{equation}%
The general multi mode noisy coherent states can be written as
\begin{eqnarray}
\rho (\{\Lambda _{k}\})_{\mathrm{mix}} &=&\prod_{k=1}^{\infty }\int
d^{2}\Gamma _{k}~\frac{1}{\mathcal{N}(\Gamma _{k})}~p(\Gamma _{k},\Lambda
_{k})\exp (\Gamma _{k}\frac{a_{k}^{\dagger }}{k})|0\rangle _{k}\langle
0|_{k}\exp (\bar{\Gamma}_{k}\frac{a_{k}}{k}).  \notag \\
&&
\end{eqnarray}

The fidelity between the multi-mode pure coherent state and noisy multi-mode
coherent state is
\begin{equation}
\langle Coh(\Lambda )\vert \rho (\{\Lambda _{k}\})_{\mathrm{mix}}\vert
Coh(\Lambda )\rangle =\prod_{k=1}^{\infty }\frac{1}{1+\frac{\mu _{k}}{k}}%
=\prod_{k=1}^{\infty }\frac{1}{1+\langle \hat{n}_{k}\rangle }.
\end{equation}

If we consider $\frac{|\Lambda _{k}|^{2}}{k}=\frac{|\Lambda |^{2k}}{k}$, the
total exitation energy $E=\langle \hat{E}\rangle $ of the noisy multi-mode
coherent state is%
\begin{equation}
\langle \hat{E}\rangle =\sum_{k}k\langle \hat{N}\rangle _{\text{\textrm{mix}}%
,k}=\sum_{k}(|\Lambda |^{2k}+\mu _{k})=\frac{|\Lambda |^{2}}{1-|\Lambda |^{2}%
}+\sum_{k}\mu _{k}.  \label{exitation energy}
\end{equation}%
The total excitation energy, in the case $\mu _{k}=0$, recovers exactly the
calculation for the pure coherent state case $\langle \hat{E}\rangle
_{Coh(\Lambda )}=\frac{|\Lambda |^{2}}{1-|\Lambda |^{2}}$~in Sec \ref{sec
2.2}.

In the limit $\mu _{k}\rightarrow 0$, $\Lambda _{k}=\Lambda ^{k}$, since the
distribution function becomes delta function,
\begin{eqnarray}
\rho (\{\Lambda ^{k}\})_{\mathrm{mix}}|_{\mu _{k}\rightarrow 0}
&=&\prod_{k=1}^{\infty }\frac{1}{\mathcal{N}(\Lambda ^{k})}\exp (\Lambda ^{k}%
\frac{a_{k}^{\dagger }}{k})\vert 0\rangle _{k}\langle 0\vert _{k}\exp (\bar{%
\Lambda}^{k}\frac{a_{k}}{k})  \notag \\
&=&\frac{1}{\mathcal{N}(\Lambda )}\vert Coh(\Lambda )\rangle \langle
Coh(\Lambda )|,
\end{eqnarray}
which recovers the multi mode pure coherent state as in (\ref{density matrix
Coh}).

We can turn on noise on one mode or on a few modes. This is related to the
partial thermalization in these modes on the subspace $\mathcal{H}_{k}$ of
the Hilbert space. In this partial thermalization, a subset of the degree of
freedom in the quantum system reaches a maximal entropy constrained by a
given amount of energy of the excitation; the gravity dual of such partial
thermalization is analyzed in \cite{Berenstein:2018lrm}. Such maximal
entropy state is related to black holes \cite{Witten:1998zw} or small black
holes, in the deconfined phase. In the confined phase, they are gas of
thermal radiations. The distinguishing between pure microstates and thermal
states have been analyzed in \cite{Balasubramanian:2007qv,Skenderis:2007yb,Mathur:2005ai,Balasubramanian:2005mg,Balasubramanian:2007zt} by methods of gauge/gravity duality and quantum
gravitational theory.

Here we used coherent state basis to analyse the superposition of density
matrices, which yields a mixed state that is nearby to the pure coherent
state, by adding noise to the pure coherent state. There is also another
basis analyzing superposition of nearby pure states \cite{Berenstein:2017abm,Berenstein:2017rrx,Koch:2008ah}, which is related to code
subspaces \cite{Berenstein:2017rrx}.

Now we analyze the single mode noisy coherent states from more perspectives
and later in relation to gauge/gravity duality. We could first make an
expansion. In each mode,%
\begin{eqnarray}
\rho (\Lambda _{k})_{\mathrm{mix}} &=&\sum_{i,j=0}^{\infty }\left( \int
d^{2}\Gamma _{k}\frac{p(\Gamma _{k},\Lambda _{k})}{\mathcal{N}(|\Gamma _{k}|)%
}(\Gamma _{k})^{i}(\bar{\Gamma}_{k})^{j}\right) \frac{1}{i!j!k^{i+j}}%
(a_{k}^{\dagger })^{i}|0\rangle _{k}\langle 0|_{k}(a_{k})^{j}.  \notag \\
&&
\end{eqnarray}%
We denote $c_{i,j}$ the density matrix elements in this basis $\frac{1}{%
\sqrt{k^{i}i!}}(a_{k}^{\dagger })^{i}|0\rangle _{k}$. The von Neumann
entropy associated to the density matrix is
\begin{equation}
s(\rho (\Lambda _{k})_{\mathrm{mix}})=-\mathrm{Tr}_{~\mathcal{H}_{k}}~\rho
(\Lambda _{k})_{\mathrm{mix}}\log \rho (\Lambda _{k})_{\mathrm{mix}}.
\end{equation}

\subsection{Relation to mixed thermal states}

\label{sec 4.2}

We can consider small $|\Lambda _{k}|/\sqrt{\mu _{k}}$ regime. The density
matrix is approximately diagonal in this regime, with diagonal matrix
elements%
\begin{equation}
c_{i,i}=\int d^{2}\Gamma _{k}\frac{e^{-|\Gamma _{k}-\Lambda _{k}|^{2}/\mu
_{k}}}{\pi \mu _{k}\mathcal{N}(|\Gamma _{k}|)}|\Gamma _{k}|^{2i}.
\end{equation}%
We can explicitely integrate this in the small $\mu _{k}$ approximation,
which is proportional to $c_{i,i}\propto |\Lambda _{k}|^{2i}\propto
e^{-\beta ki}$, where we make $\beta =-\ln |\Lambda |+\mathrm{const}$. Since
$\sum_{i}c_{i,i}=1$, we have approximately%
\begin{equation}
c_{i,i}\simeq (1-e^{-\beta k})^{-1}e^{-\beta ki}.
\end{equation}

These could be expanded in the trace basis $|t_{k}^{i}\rangle
=(a_{k}^{\dagger })^{i}|0\rangle _{k}$, $\langle \bar{t}_{k}^{j}|=\langle
0|_{k}(a_{k})^{j}$. Hence We can denote $|t_{k}^{i}\rangle =\frac{1}{(\sqrt{%
N^{k}})^{i}}($\textrm{Tr}$(Y^{k}))^{i}|0\rangle ,\langle \bar{t}_{k}^{i}|=%
\frac{1}{(\sqrt{N^{k}})^{i}}\langle 0|($\textrm{Tr}$(\bar{Y}^{k}))^{i}$ in
the context of gauge theory.$~$And hence in this approximation, we have%
\begin{equation}
\rho \simeq \frac{1}{(1-e^{-\beta k})}\sum_{i=0}^{\infty }\frac{e^{-\beta ki}%
}{k^{i}i!(\sqrt{N^{k}})^{2i}}(\mathrm{Tr}(Y^{k}))^{i}|0\rangle \langle 0|(%
\mathrm{Tr}(\bar{Y}^{k}))^{i}.
\end{equation}%
All trace numbers and $U(1)$ charges, which is also the R-charges in the
context of gauge/gravity duality, are allowed, hence it is in a grand
canonical ensemble. It is reminiscent to the ensemble of microstates of
one-charge black hole in AdS. In the confined phase, this can also be viewed
as a thermal gas of multi gravitons. While, in the deconfined phase, they
could be related to black holes \cite{Witten:1998zw,Balasubramanian:2007bs,Fareghbal:2008ar} or small black holes.

\section{Coherent states of Young tableaux}

\label{sec 5} \renewcommand{\theequation}{5.\arabic{equation}} %
\setcounter{equation}{0} \renewcommand{\thethm}{5.\arabic{thm}} %
\setcounter{thm}{0} \renewcommand{\theprop}{5.\arabic{prop}} %
\setcounter{prop}{0}

In this general section \ref{sec 5}, we consider another class of coherent
states, different from the class in Sec \ref{sec 2}. The class in Sec \ref%
{sec 2} could be viewed as coherent states of multi traces. This class is
for coherent states of multi columns of Young tableaux (YT).

\subsection{Coherent states of single-row Young tableaux}

\label{sec 5.1}

We denote $\Delta _{n}$ as a Young tableau with a single row of length $n$.
Consider the creation and annihilation operators $A^{\dagger },A$ on the
single-row Young tableau states $|\Delta _{n}\rangle $
\begin{equation}
A^{\dagger }|\Delta _{n}\rangle =\sqrt{n+1}|\Delta _{n+1}\rangle ,\quad
\quad \quad A|\Delta _{n}\rangle =\sqrt{n}|\Delta _{n-1}\rangle ,
\label{crea. and annih. 01}
\end{equation}%
with $[A,A^{\dagger }]=1$ and $A|\Delta _{0}\rangle =0$.~The first equation
in (\ref{crea. and annih. 01}) can also be written as%
\begin{equation}
A^{\dagger }|\Delta _{n-1}\rangle =\sqrt{n}|\Delta _{n}\rangle .~
\end{equation}%
Hence $|\Delta _{n}\rangle =\frac{(A^{\dagger })^{n}}{\sqrt{n!}}|\Delta
_{0}\rangle $ and $\Vert \Delta _{n}\Vert =1$. The action of $A^{\dagger }$
is adding one column on the Young tableau. The action of $A$ is removing one
column from the Young tableau. This is in the large $N$ limit.

We can construct the coherent states of Young tableaux,%
\begin{equation}
|\Lambda \rangle =e^{-\frac{|\Lambda |^{2}}{2}}e^{\Lambda A^{\dagger
}}|\Delta _{0}\rangle =e^{-\frac{|\Lambda |^{2}}{2}}\sum\limits_{n=0}^{%
\infty }\frac{1}{\sqrt{n!}}\Lambda ^{n}|\Delta _{n}\rangle .  \label{coh. YT}
\end{equation}%
We have $A|\Lambda \rangle =\Lambda |\Lambda \rangle $ and $\Vert |\Lambda
\rangle \Vert =1$. Here the range of $\Lambda $ is $0<|\Lambda |<\infty $.
Note that this state (\ref{coh. YT})~is different from $|Coh(\Lambda
)\rangle $ because of the coefficients in the expansion. The $|\Lambda
\rangle $ and the YT state $|\Delta _{n}\rangle $ are entangled states in
the tensor product of the multi-trace Hilbert spaces $\otimes _{k}\mathcal{H}%
_{k}$ in Sec. \ref{sec 2}.

The inner product is%
\begin{equation}
\langle \Lambda |\Lambda ^{\prime }\rangle =e^{-\frac{1}{2}|\Lambda |^{2}-%
\frac{1}{2}|\Lambda ^{\prime }|^{2}+\bar{\Lambda}\Lambda ^{\prime
}},~~~~~~~|\langle \Lambda |\Lambda ^{\prime }\rangle |~=e^{-\frac{1}{2}%
|\Lambda -\Lambda ^{\prime }|^{2}}.
\end{equation}

The number operator is $\hat{N}=A^{\dagger }A$. The excitation energy
operator is $\hat{E}=A^{\dagger }A$, with the zero point ground state energy
subtracted out in the definition. One box of Young tableau has one unit of
energy. The phase shift operator is $\exp (i\theta \hat{N})=\exp (i\theta
A^{\dagger }A)=\exp (i\theta \hat{E})$, and the action is $\exp (i\theta
\hat{E})|\Lambda \rangle =|\Lambda e^{i\theta }\rangle $. The parity
operator is $\hat{P}=\exp (i\pi \hat{E})$.

The generating function is%
\begin{equation}
\langle \exp (i\theta \hat{E})\rangle _{|\Lambda \rangle }=e^{|\Lambda
|^{2}(e^{i\theta }-1)}.
\end{equation}%
Hence $\langle \hat{E}\rangle _{|\Lambda \rangle }=|\Lambda |^{2}$ and $%
(\Delta E)_{|\Lambda \rangle }^{2}=|\Lambda |^{2}$.

\subsection{Coherent states of $k$-row Young tableaux}

\label{sec 5.2}

We can define coherent states of $k$-row Young tableaux. We denote $\Delta
_{n,k}$ as a Young tableau with $n$ columns each with a column-length $k$.
In the context of gauge theory, this state can be written as a Schur
polynomial operator labelled by a Young tableau $\Delta _{n,k}~$with $n$
columns and $k$ rows \cite{Corley:2001zk}.

The $\Delta _{n,k}$ is a Young tableau with $n$ columns each with a
column-length $k$. Consider the creation and annihilation operators $%
A_{k}^{\dagger },A_{k}$ on the multi-row Young tableau states $|\Delta
_{n,k}\rangle $,
\begin{equation}
A_{k}^{\dagger }|\Delta _{n,k}\rangle =\sqrt{k(n+1)}|\Delta _{n+1,k}\rangle
,\quad \quad \quad A_{k}|\Delta _{n,k}\rangle =\sqrt{kn}|\Delta
_{n-1,k}\rangle ,  \label{crea. and annih. 02}
\end{equation}%
with $\frac{1}{k}[A_{k},A_{k}^{\dagger }]=1$ and $A_{k}|\Delta _{0,k}\rangle
=0$.~The first equation in (\ref{crea. and annih. 02}) can also be written as%
\begin{equation}
A_{k}^{\dagger }|\Delta _{n-1,k}\rangle =\sqrt{kn}|\Delta _{n,k}\rangle .
\end{equation}%
In other words,$~\frac{1}{\sqrt{k}}A_{k}^{\dagger }$ and $\frac{1}{\sqrt{k}}%
A_{k}~$play the role of ordinary creation and annihilation operators, with
the $\frac{1}{\sqrt{k}}$ factor due to our particular convention of the
definition. This is in the large $N$ limit. Hence $|\Delta _{n,k}\rangle =%
\frac{(A_{k}^{\dagger })^{n}}{\sqrt{k^{n}n!}}|\Delta _{0,k}\rangle $ and $%
\Vert \Delta _{n,k}\Vert =1$. The action of $A_{k}^{\dagger }$ is adding one
column of length-$k$ on the Young tableau. The action of $A_{k}$ is removing
one column of length-$k$ from the Young tableau. The YT state $|\Delta
_{n,k}\rangle $ is an entangled state in the tensor product of the
multi-trace Hilbert spaces $\otimes _{k}\mathcal{H}_{k}$ in Sec. \ref{sec 2}%
. These states contain nontrivial entanglement stored between different
multi-trace Hilbert spaces $\mathcal{H}_{k}$. See detailed analysis of this
statement in \cite{Lin:2017dnz}. The entanglement entropy of these states,
entangled in $\otimes _{k}\mathcal{H}_{k}$, were computed in \cite%
{Lin:2017dnz,Berenstein:2017abm}.

The action of the excitation energy operator $\hat{E}=\sum_{k}\hat{E}_{k}$
and the phase shift operator $\exp (i\theta \hat{E})$ on the YT states are
as follows: $\hat{E}|\Delta _{n,k}\rangle =kn|\Delta _{n,k}\rangle $, $\exp
(i\theta \hat{E})|\Delta _{n,k}\rangle =e^{ikn\theta }|\Delta _{n,k}\rangle $%
, and $\langle \exp (i\theta \hat{E})\rangle _{|\Delta _{n,k}\rangle
}=e^{ikn\theta }$. We also have $\hat{P}_{k}|\Delta _{n,k}\rangle
=(-1)^{n}|\Delta _{n,k}\rangle $ and $\exp (i\pi \hat{E})|\Delta
_{n,k}\rangle $ $=(-1)^{kn}|\Delta _{n,k}\rangle $. The phase rotation of
the matrix field induces a phase rotation for each box of the Young tableau.

This class of coherent states are also described in \cite%
{Balasubramanian:2018yjq,Dhar:2005fg}, and our case is a special case
of them. We can construct the coherent states of multi-row ($k$-row) Young
tableaux,%
\begin{equation}
|\Lambda _{k}\rangle _{k}=e^{-\frac{|\Lambda _{k}|^{2}}{2k}}e^{\frac{1}{k}%
\Lambda _{k}A_{k}^{\dagger }}|\Delta _{0,k}\rangle =e^{-\frac{|\Lambda
_{k}|^{2}}{2k}}\sum\limits_{n=0}^{\infty }\frac{1}{\sqrt{k^{n}n!}}(\Lambda
_{k})^{n}|\Delta _{n,k}\rangle .  \label{coh. YT 02}
\end{equation}%
The subscript $k$ means `mode' $k$. We have $A_{k}|\Lambda _{k}\rangle
_{k}=\Lambda _{k}|\Lambda _{k}\rangle _{k}$ and $\Vert |\Lambda _{k}\rangle
_{k}\Vert =1$. Here the range of $\Lambda _{k}$ is $0<|\Lambda _{k}|<\infty $%
. The amplitude of the coherent states, in the conventional notation of
quantum optics, is $\frac{\Lambda _{k}}{\sqrt{k}}$. We sometimes also denote
$\frac{\Lambda _{k}}{\sqrt{k}}=z_{k}$. In the expansion of this coherent
state, the $|\Delta _{n,k}\rangle $ and $|\Delta _{0,k}\rangle $ are the
analogs of the photon states $|n\rangle $ and $|0\rangle $ in quantum optics
and quantum information theory.

We can refer to the giant gravitons wrapping internal sphere directions, as
sphere giant gravitons, and those wrapping AdS directions, as dual giant
gravitons. Each column is a sphere giant graviton with momentum $k$. This
state (\ref{coh. YT 02}) is a coherent state of length-$k~$columns, and in
other words a coherent state of momentum-$k~$sphere giant gravitons, with an
average sphere giant graviton number $\langle \hat{E}\rangle /k$. The
variance of the sphere giant graviton number is $(\Delta E)^{2}/k^{2}$. It
has a dual interpretation that it is also a state of $k$ dual giant
gravitons with a fluctuating size and an average size $\langle \hat{E}%
\rangle /k$. The variance of the size of the dual giant gravitons is $%
(\Delta E)^{2}/k^{2}=\mathrm{QFI}/(4k^{2})$, which is related to the quantum
Fisher information.

There is a relation between YT states and fermions. The $i$-th fermion
excitation energy is $E_{i}=c_{i}+i,$ where $i$ is the $i$-th row and $c_{i}$
is the row-length of the $i$-th row. Here, the average row-length is $%
\langle \hat{E}\rangle /k$. So the average fermion excitation energy of the
above coherent state is $\langle \hat{E}\rangle /k+N/2$. The state $|\Lambda
_{k}\rangle _{k}$ can also be viewed as a $k$-fermion droplet or blob
centered around $\Lambda _{k}/\sqrt{k}$, with $k$ fermions in that droplet.
This state has very good localizability in phase space. The special case of
single-row, in the preceding section, corresponds to $k=1,\Lambda =$ $%
\Lambda _{1}$.

Define $\hat{x}_{k}=\frac{1}{\sqrt{2k}}(A_{k}+A_{k}^{\dagger })$, $\hat{p}%
_{k}=\frac{1}{\sqrt{2k}i}(A_{k}-A_{k}^{\dagger })$, then $\langle \hat{x}%
_{k}\rangle =\frac{1}{\sqrt{k}}\mathrm{Re}{~}\Lambda _{k},~\langle \hat{p}%
_{k}\rangle =\frac{1}{\sqrt{k}}\mathrm{Im}{~}\Lambda _{k}$. Hence the state$%
~|\Lambda _{k}\rangle _{k}$ is a blob or a droplet centered around $(\frac{1%
}{\sqrt{k}}\mathrm{Re}{~}\Lambda _{k},\frac{1}{\sqrt{k}}\mathrm{Im}{~}%
\Lambda _{k})$ in the phase space, and subject to the Heisenberg uncertainty
principle.

The inner product is%
\begin{equation}
\langle \Lambda _{k}|\Lambda _{k}^{\prime }\rangle _{k}=e^{-\frac{1}{2k}%
|\Lambda _{k}|^{2}-\frac{1}{2k}|\Lambda _{k}^{\prime }|^{2}+\frac{1}{k}\bar{%
\Lambda}_{k}\Lambda _{k}^{\prime }},~~~~~~~|\langle \Lambda _{k}|\Lambda
_{k}^{\prime }\rangle |~=e^{-\frac{1}{2k}|\Lambda _{k}-\Lambda _{k}^{\prime
}|^{2}}.
\end{equation}

The number operator is $\hat{N}_{k}=\frac{1}{k}A_{k}^{\dagger }A_{k}$. The
excitation energy operator is $\hat{E}_{k}=A_{k}^{\dagger }A_{k}$, with the
zero point ground state energy subtracted out in the definition. One box of
Young tableau has one unit of energy. We have that $\theta _{k}=k\theta $.
The phase shift operator is $\exp (i\theta _{k}\hat{N}_{k})=\exp (i\frac{%
\theta _{k}}{k}\hat{E}_{k})$. The action is%
\begin{equation}
\exp (i\frac{\theta _{k}}{k}\hat{E}_{k})|\Lambda _{k}\rangle _{k}=|\Lambda
_{k}e^{i\theta _{k}}\rangle _{k}.
\end{equation}%
The parity operator for the $k$-row state is $\hat{P}_{k}=\exp (i\frac{\pi }{%
k}\hat{E}_{k})$, and the action is $\hat{P}_{k}|\Lambda _{k}\rangle
_{k}=\lvert -\Lambda _{k}\rangle _{k}$.

The generating function is%
\begin{equation}
\langle \exp (i\frac{\theta _{k}}{k}\hat{E}_{k})\rangle _{|\Lambda
_{k}\rangle _{k}}=\langle \exp (i\theta \hat{E})\rangle _{|\Lambda
_{k}\rangle _{k}}=e^{\frac{|\Lambda _{k}|^{2}}{k}(e^{i\theta _{k}}-1)}.
\end{equation}%
Hence $\langle \hat{E}\rangle _{|\Lambda _{k}\rangle _{k}}=|\Lambda
_{k}|^{2} $, $\langle \hat{N}_{k}\rangle _{|\Lambda _{k}\rangle _{k}}=\frac{%
|\Lambda _{k}|^{2}}{k}=|z_{k}|^{2}$, and $(\Delta E)_{|\Lambda _{k}\rangle
_{k}}^{2}=k|\Lambda _{k}|^{2}$, $(\Delta N)_{|\Lambda _{k}\rangle _{k}}^{2}=%
\frac{|\Lambda _{k}|^{2}}{k}=\langle \hat{N}_{k}\rangle _{|\Lambda
_{k}\rangle _{k}}$.

In the context of gauge/string duality, $\lvert \Delta _{n,k}\rangle ~$and $%
\lvert \Delta _{n}\rangle =\lvert \Delta _{n,1}\rangle $ is also a D-brane
state, and more specifically, a bound state of $k$ dual giant gravitons with
size $n$. In the geometric dual, the state$\lvert \Delta _{n,k}\rangle$ is
also a black ring with area $k$ located around the radial location $n$ in
the phase space plane. These interpretations are used in Sec. \ref{sec 7}.

This class of coherent states is similar to the first type of coherent
states discussed in \ref{sec 2.1}. The two different types differ as the
difference between the multi trace case and the multi column case. Because
the Young tableau has expansion in terms of products of multi-traces, this
coherent state $|\Lambda _{k}\rangle _{k}$ for fixed $k$ is at the level of
`multi-mode' in the multi-trace basis, like the multi-mode $\lvert
Coh(\Lambda )\rangle $ in Sec. \ref{sec 2}.

\subsection{Multi mode coherent states of Young tableaux}

Now we denote $\mathcal{H}_{k}$ as the Hilbert space of all column length $k$
tableaux, an analog of `mode' $k$, and $k\in \mathbb{Z}_{>0}$. The $|\Delta
_{0,k}\rangle =|0\rangle $ is the vacuum state. Consider $\mathcal{H}$ $=%
\mathcal{H}_{1}\otimes \mathcal{H}_{2}\otimes \dots =\otimes _{k}\mathcal{H}%
_{k}$. The `multi-mode' coherent states of YT is
\begin{eqnarray}
\Psi (\{\Lambda _{k}\}) &=&\prod_{k}|\Lambda _{k}\rangle _{k}  \notag \\
&=&\exp (-\sum_{k}\frac{|\Lambda _{k}|^{2}}{2k})\exp (\sum_{k}\frac{1}{k}%
\Lambda _{k}A_{k}^{\dagger })|0\rangle .
\end{eqnarray}%
These are coherent states involving arbitrary-shape Young tableaux, with
allowed shapes \cite{Fulton,Sagan,Fulton Harris}. The $k$
denotes and labels the column length. If we include all possible $k$ and all
possible shapes, each individual configuration is very reminiscent of a
microstate configuration of a fuzzball \cite{Mathur:2005ai}.

For multi-mode, we define phase shift operator as $\exp (i\sum\limits_{k}%
\frac{\theta _{k}}{k}\hat{E}_{k})$. We consider$~\frac{\theta _{k}}{k}%
=\theta $. Hence%
\begin{equation}
\exp (i\sum\limits_{k}\frac{\theta _{k}}{k}\hat{E}_{k})=\exp (i\theta \hat{E}%
),
\end{equation}%
where $\hat{E}_{k}=A_{k}^{\dagger }A_{k}~$and $\hat{E}=\sum\limits_{k}\hat{E}%
_{k}$. The action of the phase shift operator is%
\begin{equation}
\exp (i\theta \hat{E})|\Lambda _{k}\rangle _{k}=|\Lambda _{k}e^{ik\theta
}\rangle _{k}.
\end{equation}

The symplectic form in the configuration space of the droplets can be
written \cite{Grant:2005qc,Balasubramanian:2018yjq} as%
\begin{equation}
\Omega =\sum_{k}\frac{1}{k}\delta \Lambda _{k}\wedge \delta \bar{\Lambda}%
_{k},
\end{equation}%
where $k$ indexes column lengthes. The $\frac{1}{k}$ factor is due to our
convention, since $\frac{1}{\sqrt{k}}\Lambda _{k}$ is the amplitude or the
complex parameter of the coherent states in the phase space. This captures
the deformation modes of the droplet geometries, and hence plays important
role in the deformation space of microstate geometries, and also has
implication for the stretched horizon of the superstar geometry \cite%
{Balasubramanian:2018yjq}.

For example, the `two-mode' coherent states of YT is
\begin{equation}
\Psi =e^{-\frac{|\Lambda _{k_{1}}|^{2}}{2k_{1}}}e^{\frac{1}{k_{1}}\Lambda
_{k_{1}}A_{k_{1}}^{\dagger }}e^{-\frac{|\Lambda _{k_{2}}|^{2}}{2k_{2}}}e^{%
\frac{1}{k_{2}}\Lambda _{k_{2}}A_{k_{2}}^{\dagger }}|0\rangle .
\end{equation}

Here we use the column number to classify the `modes'. We can alternatively
use the corners of YT to classify the `modes', e.g. \cite{Berenstein:2017rrx,Koch:2008ah}. The former is more convenient for creating D-brane
state, while the later is more convenient for creating closed string state.
See also other closely related basis describing different degrees of freedom
using Young tableaux \cite{deMelloKoch:2020jmf,Ramgoolam:2018ceu,Lin:2017vfn,Balasubramanian:2018yjq,Dhar:2005su}.

\subsection{Cat states}

Similar to Sec. \ref{sec 3} where we consider the Schrodinger cat state for
the first type of coherent state, in this section we analyze the cat state
for the second type of coherent state. For a general $k$, we have the cat
state
\begin{eqnarray}
|cat_{\pm }(\Lambda _{k})\rangle _{k} &=&\frac{1}{\sqrt{N_{k,\pm }}}%
(|\Lambda _{k}\rangle _{k}\pm |-\Lambda _{k}\rangle _{k}) \\
&=&\frac{1}{\sqrt{N_{k,\pm }}}(1\pm \hat{P}_{k})|\Lambda _{k}\rangle _{k},
\end{eqnarray}%
where $|cat_{\pm }(\Lambda _{k})\rangle _{k}$ has unit norm and here $%
N_{k,\pm }=2(1\pm e^{-\frac{2}{k}|\Lambda _{k}|^{2}})$. This is a
superposition of two blobs. They are centered at $(\frac{1}{\sqrt{k}}\mathrm{%
Re}{~}\Lambda _{k},\frac{1}{\sqrt{k}}\mathrm{Im}{~}\Lambda _{k})$ and $(-%
\frac{1}{\sqrt{k}}\mathrm{Re}{~}\Lambda _{k},-\frac{1}{\sqrt{k}}\mathrm{Im}{~%
}\Lambda _{k})$ respectively in the phase space plane. This phase space
plane is the droplet plane in the gravity side. The larger the$~\frac{1}{k}%
|\Lambda _{k}|^{2}$, the more far apart they are. We have $\langle \hat{P}%
_{k}\rangle _{cat_{\pm }(\Lambda _{k})}=\pm 1$,$~(\Delta P_{k})_{cat_{\pm
}(\Lambda _{k})}^{2}=0$. These states can be used to construct qubits.

The generating function is%
\begin{eqnarray}
\langle \exp (i\frac{\theta _{k}}{k}\hat{E}_{k})\rangle _{|cat_{\pm
}(\Lambda _{k})\rangle _{k}} &=&\frac{e^{-\frac{1}{k}|\Lambda _{k}|^{2}}}{%
(1\pm e^{-\frac{2}{k}|\Lambda _{k}|^{2}})}[\exp (\frac{1}{k}|\Lambda
_{k}|^{2}e^{i\theta _{k}})\pm \exp (-\frac{1}{k}|\Lambda _{k}|^{2}e^{i\theta
_{k}})].  \notag \\
&&
\end{eqnarray}%
We have that:%
\begin{eqnarray}
\langle \hat{E}\rangle _{cat_{+}(\Lambda _{k})} &=&|\Lambda _{k}|^{2}\mathrm{%
\tanh }(\frac{|\Lambda _{k}|^{2}}{k}),~~\langle \hat{E}\rangle
_{cat_{-}(\Lambda _{k})}=|\Lambda _{k}|^{2}\mathrm{\tanh }(\frac{|\Lambda
_{k}|^{2}}{k})^{-1}, \\
(\Delta E)_{cat_{+}(\Lambda _{k})}^{2} &=&k|\Lambda _{k}|^{2}\mathrm{\tanh }(%
\frac{|\Lambda _{k}|^{2}}{k})+|\Lambda _{k}|^{4}\cosh (\frac{|\Lambda
_{k}|^{2}}{k})^{-2}, \\
(\Delta E)_{cat_{-}(\Lambda _{k})}^{2} &=&k|\Lambda _{k}|^{2}\mathrm{\tanh }(%
\frac{|\Lambda _{k}|^{2}}{k})^{-1}-|\Lambda _{k}|^{4}\sinh (\frac{|\Lambda
_{k}|^{2}}{k})^{-2}.
\end{eqnarray}

For big $|\Lambda _{k}|\gg 1$, we have that $\langle \hat{E}\rangle
_{cat_{\pm }(\Lambda _{k})}\simeq |\Lambda _{k}|^{2}\pm 2|\Lambda
_{k}|^{2}e^{-\frac{2|\Lambda _{k}|^{2}}{k}}$, and$~(\Delta E)_{cat_{\pm
}(\Lambda _{k})}^{2}\simeq k|\Lambda _{k}|^{2}\pm 4|\Lambda _{k}|^{4}e^{-%
\frac{2|\Lambda _{k}|^{2}}{k}}$. For big $|\Lambda _{k}|\gg 1,\frac{\Delta E%
}{\langle \hat{E}\rangle }|_{cat_{\pm }(\Lambda _{k})}\simeq \frac{\sqrt{k}}{%
|\Lambda _{k}|}$. Hence, for relatively big coherent state amplitude, the $%
cat_{+},cat_{-}$ are more close to the coherent state.

Moreover, it has been known that the expression of the cat state can be
approximated by squeezed state:
\begin{eqnarray}
|cat_{+}(\Lambda _{k})\rangle _{k} &\sim &S(\xi )|\Delta _{0,k}\rangle , \\
|cat_{-}(\Lambda _{k})\rangle _{k} &\sim &S(\xi )|\Delta _{1,k}\rangle ,
\end{eqnarray}%
where $S(\xi )=e^{-\frac{\xi }{k}(A_{k}^{\dagger }A_{k}^{\dagger
}-A_{k}A_{k})}$, with high fidelity \cite{Nielsen etal,Andersen etal}%
, and we have generalized them to new states defined using Young tableaux,
from those happens in quantum information theory \cite{Nielsen etal,Andersen etal}. This approximation is practically useful, since certain
calculation can be simplified with it.

\subsection{Noisy coherent states of multi-row Young tableaux}

The preceding sec. \ref{sec 4.1} is on a class of noisy coherent states of
multi traces. In this section, we discuss a different class, which is a
class of noisy coherent states of Young tableaux. We define $%
A_{k}=d_{k}+\Lambda _{k}$, $A_{k}^{\dagger }=d_{k}^{\dagger }+\bar{\Lambda}%
_{k}$, and $\frac{1}{k}[d_{k},~d_{k}^{\dagger }]=1$. Hence,%
\begin{eqnarray}
d_{k}|\Lambda _{k}\rangle _{k} &=&0,~~~ \\
d_{k}|\Gamma _{k}\rangle _{k} &=&(\Gamma _{k}-\Lambda _{k})|\Gamma
_{k}\rangle .~
\end{eqnarray}%
We have $\langle \Gamma _{k}|(d_{k}^{\dagger })^{l}(d_{k})^{l}|\Gamma
_{k}\rangle _{k}=|\Gamma _{k}-\Lambda _{k}|^{2l}$.$~$The thermal particle
number operator is $\hat{n}_{k}=\frac{1}{k}d_{k}^{\dagger }d_{k}$.

Consider a mixed state density matrix,
\begin{eqnarray}
\rho (\Lambda _{k})_{\mathrm{mix}} &=&\sum_{l=0}^{\infty }\frac{\langle \hat{%
n}_{k}\rangle ^{l}}{(1+\langle \hat{n}_{k}\rangle )^{l+1}}(d_{k}^{\dagger
})^{l}|\Lambda _{k}\rangle _{k}\langle \Lambda _{k}|_{k}(d_{k})^{l}
\label{distribution 02} \\
&=&\int d^{2}\Gamma _{k}\frac{1}{\pi k\langle \hat{n}_{k}\rangle }\exp (-%
\frac{|\Gamma _{k}-\Lambda _{k}|^{2}}{k\langle \hat{n}_{k}\rangle })|\Gamma
_{k}\rangle _{k}\langle \Gamma _{k}|_{k}.
\end{eqnarray}%
The distribution $p(l)=\frac{\langle \hat{n}_{k}\rangle ^{l}}{(1+\langle
\hat{n}_{k}\rangle )^{l+1}}$ in (\ref{distribution 02}) is Bose-Einstein
distribution of random columns on top of a coherent state. The random
columns are the noise.

Hence, the density matrix of noisy coherent states of Young tableaux is
\begin{eqnarray}
\rho _{\text{\textrm{mix}}}(\Lambda _{k}) &=&\int d^{2}\Gamma _{k}~p(\Gamma
_{k},\Lambda _{k})\frac{1}{\mathcal{N}(\Gamma _{k})}\exp (\frac{1}{k}\Gamma
_{k}A_{k}^{\dagger })|\Delta _{0,k}\rangle \langle \Delta _{0,k}|\exp (\frac{%
1}{k}\bar{\Gamma}_{k}A_{k}),  \notag \\
p(\Gamma _{k},\Lambda _{k}) &=&\frac{1}{\pi \mu _{k}}\exp (-|\Gamma
_{k}-\Lambda _{k}|^{2}/\mu _{k}).
\end{eqnarray}

The fidelity between the pure coherent state $\rho _{\mathrm{pure}}=\rho
_{|\Lambda _{k}\rangle _{k}}$ and noisy coherent state $\rho _{\mathrm{mix}%
}=\rho (\Lambda _{k})_{\mathrm{mix}}$ is%
\begin{equation}
\text{\textrm{tr}}(\rho _{\mathrm{mix}}\rho _{\mathrm{pure}})=\langle
\Lambda _{k}|\rho (\Lambda _{k})_{\mathrm{mix}}|\Lambda _{k}\rangle _{k}=%
\frac{1}{1+\frac{\mu _{k}}{k}}=\frac{1}{1+\langle \hat{n}\rangle _{k}}.
\end{equation}

The total particle number operator in mode $k$ is $\hat{N}_{k}=\frac{1}{k}%
A_{k}^{\dagger }A_{k}$. In the mixed state,%
\begin{eqnarray}
\langle \hat{N}_{k}\rangle _{\text{\textrm{mix}}} &=&\mathrm{tr~}(\rho _{%
\text{\textrm{mix}}}(\Lambda _{k})\frac{1}{k}A_{k}^{\dagger }A_{k})=\int
d^{2}\Gamma _{k}\frac{1}{\pi \mu _{k}}\exp (-|\Gamma _{k}-\Lambda
_{k}|^{2}/\mu _{k})~|\Gamma _{k}|^{2}  \notag \\
&=&\frac{1}{k}|\Lambda _{k}|^{2}+\frac{1}{k}\mu _{k}=\langle \hat{N}%
_{k}\rangle _{\text{\textrm{pure}}}+\langle \hat{n}_{k}\rangle .
\end{eqnarray}%
The generating function for the mixed state is $\langle \exp (i\frac{\theta
_{k}}{k}\hat{E}_{k})\rangle _{\rho _{\mathrm{mix}}}=~$\textrm{tr}$(e^{i\frac{%
\theta _{k}}{k}\hat{E}_{k}}\rho _{\mathrm{mix}})$. The excitation energy is%
\begin{eqnarray}
\langle \hat{E}_{k}\rangle &=&k\langle \hat{N}_{k}\rangle =|\Lambda
_{k}|^{2}+k\langle \hat{n}_{k}\rangle =|\Lambda _{k}|^{2}+\mu _{k}. \\
(\Delta E_{k})_{~}^{2} &=&k\langle \hat{E}_{k}\rangle +2|\Lambda
_{k}|^{2}\mu _{k}+\mu _{k}^{2}.
\end{eqnarray}%
Of course, the $k=1$ case, is a special case in the preceding section \ref%
{sec 5.1}.

The average \emph{size} of the dual giant gravitons is $\langle \hat{E}%
\rangle /k=\frac{1}{k}|\Lambda _{k}|^{2}+\frac{1}{k}\mu _{k}$. The variance
of the size of the dual giant gravitons is $(\Delta E)^{2}/k^{2}=\langle
\hat{E}\rangle /k+\frac{2}{k^{2}}|\Lambda _{k}|^{2}\mu _{k}+\frac{1}{k^{2}}%
\mu _{k}^{2}$. Note that the shift $|\Lambda _{k}|^{2}\rightarrow |\Lambda
_{k}|^{2}+\mu _{k}~$in the size of the dual giant graviton, from a pure
state to a mixed state, is very reminiscent to the shift from a black disk
to a gray disk corresponding to the superstar geometry which may be viewed
as a mixed state of dual giant gravitons \cite{Balasubramanian:2007qv,Balasubramanian:2005mg}. There is a giant/dual-giant duality between sphere
giants and dual giants, and hence the $\langle \hat{E}\rangle /k$ is also
the \emph{number} of the sphere giant gravitons. The notation $\langle \hat{N%
}\rangle =\langle \hat{E}\rangle /k$ sphere giants here is the same as the
notation $N_{c}$ giants in \cite{Balasubramanian:2005mg,Simon:2018laf}%
.

In the small $|\Lambda _{k}|/\sqrt{\mu _{k}}$ regime,$~$we have a limit to
mixed thermal state which is approximately diagonal. Similar to Sec. \ref%
{sec 4.2}, we have then the resulting density matrix
\begin{equation}
\rho \simeq (1-e^{-\beta k})^{-1}\sum_{n=0}^{\infty }e^{-\beta kn}|\Delta
_{n,k}\rangle \langle \Delta _{n,k}\vert .  \label{density matrix 04}
\end{equation}

\subsection{Relation to mixed thermal state and its purification}

The above mixed density matrix (\ref{density matrix 04}) have a purification
in doubled Hilbert space. Consider a \emph{pure} state $\vert g\rangle ~$in$~%
\mathcal{H}_{k}^{L}\otimes \mathcal{H}_{k}^{R},$%
\begin{eqnarray}
\vert g\rangle &=&\frac{1}{\sqrt{\mathcal{N}}}\sum\limits_{n=0}^{\infty }e^{-%
\frac{\beta }{2}kn}|\Delta _{n,k}\rangle _{L}|\Delta _{n,k}\rangle _{R}
\notag \\
&=&\frac{1}{\sqrt{\mathcal{N}}}\sum\limits_{n=0}^{\infty }\frac{e^{-\frac{%
\beta }{2}kn}}{k^{n}n!}(A_{k;L}^{\dagger })^{n}(A_{k;R}^{\dagger })^{n}\vert
0\rangle _{k;L}\vert 0\rangle _{k;R},  \label{g 01}
\end{eqnarray}%
where $\mathcal{N}=(1-e^{-\beta k})^{-1}$. Here $|\Delta _{n,k}\rangle
_{L}\in \mathcal{H}_{k}^{L}\subset \mathcal{H}^{L}$ and $|\Delta
_{n,k}\rangle _{R}\in \mathcal{H}_{k}^{R}\subset \mathcal{H}^{R}$.

This state can be derived from a squeezed state,
\begin{equation}
|g\rangle =\frac{1}{\sqrt{\mathcal{N}}}e^{-\frac{\xi }{k}(A_{k;L}^{\dagger
}A_{k;R}^{\dagger }-A_{k;L}A_{k;R})}|0\rangle _{L}|0\rangle _{R},~~~~~\tanh
\xi =-e^{-\frac{\beta }{2}k}.  \label{g}
\end{equation}%
Here $S(\xi )=e^{-\frac{\xi }{k}(A_{k;L}^{\dagger }A_{k;R}^{\dagger
}-A_{k;L}A_{k;R})}$ is the squeeze operator. The $\frac{1}{k}$ factor is due
to our convention of the definition. The normalization factor is $\mathcal{N}%
=(1-\tanh ^{2}\xi )^{-1}=(1-e^{-\beta k})^{-1}$. Here $1/T=\beta =\frac{2}{k}%
\ln |\tanh \xi |^{-1}\geq 0$. In relating (\ref{g}) and (\ref{g 01}), we use
a formula
\begin{eqnarray}
&&e^{-\frac{\xi }{k}(A_{k;L}^{\dagger }A_{k;R}^{\dagger
}-A_{k;L}A_{k;R})}|0\rangle _{L}|0\rangle _{R}  \notag \\
&=&e^{-\frac{1}{k}\tanh \xi ~A_{k;L}^{\dagger }A_{k;R}^{\dagger }}(\cosh \xi
)^{-\frac{1}{k}A_{k;L}^{\dagger }A_{k;L}-\frac{1}{k}A_{k;R}^{\dagger
}A_{k;R}-1}e^{\frac{1}{k}\tanh \xi ~A_{k;L}A_{k;R}}|0\rangle _{L}|0\rangle
_{R}.  \notag \\
&&
\end{eqnarray}%
We can view $e^{iH_{\mathrm{int}}}=e^{-\frac{\xi }{k}(A_{k;L}^{\dagger
}A_{k;R}^{\dagger }-A_{k;L}A_{k;R})}$, where the interacting Hamiltonian
coupling the two copies of the Hilbert space is $H_{\mathrm{int}}=i\frac{\xi
}{k}A_{k;L}^{\dagger }A_{k;R}^{\dagger }-i\frac{\xi }{k}A_{k;L}A_{k;R}$.

Its density matrix on $\mathcal{H}_{k}^{L}\otimes \mathcal{H}_{k}^{R}$ is $%
\vert g\rangle \langle g\vert $. The partial trace gives a mixed density
matrix in $\mathcal{H}_{k}^{L},$%
\begin{eqnarray}
\rho _{L} &=&\rho _{\mathrm{mix}}~=~\mathrm{tr}_{\mathcal{H}_{k}^{R}}~\vert
g\rangle \langle g\vert  \notag \\
&=&(1-e^{-\beta k})^{-1}\sum_{n=0}^{\infty }\frac{e^{-\beta kn}}{k^{n}n!}%
(A_{k}^{\dagger })^{n}\vert 0\rangle _{k}\langle 0\vert _{k}(A_{k})^{n}
\notag \\
&=&(1-e^{-\beta k})^{-1}\sum_{n=0}^{\infty }e^{-\beta kn}|\Delta
_{n,k}\rangle \langle \Delta _{n,k}\vert ,
\end{eqnarray}%
which is the same as (\ref{density matrix 04}), hence $\vert g\rangle $ is a
purification of (\ref{density matrix 04}). This also describes thermal
sphere giant gravitons with size $k$. This can be viewed as a reduced
density matrix in the subspace of the Hilbert space, labeled by the `mode' $%
k $. This can also be viewed as being from the noisy coherent state in the
large noise limit. In the gravity dual, in the confined phase, they can be
viewed as a gas of thermal giant gravitons \cite{Armas:2012bk,
Balasubramanian:2005mg}.

\section{Quantum state discrimination and quantum detection}

\label{sec 6} \renewcommand{\theequation}{6.\arabic{equation}} %
\setcounter{equation}{0} \renewcommand{\thethm}{6.\arabic{thm}} %
\setcounter{thm}{0} \renewcommand{\theprop}{6.\arabic{prop}} %
\setcounter{prop}{0}

Imagining a measurement experiment, that Alice send signals of two quantum
states $\rho _{0},\rho _{1}$, which are linearly independent, but may or may
not be orthogonal. She sends them with a priori probabilities $\eta
_{0},\eta _{1}$, respectively. Bob choose projectors $\Pi _{0},\Pi _{1}$ to
measure these two states respectively. For each measurement, Bob has to
decide which state he has measured. There is a probability that his decision
is not correct. Bob wants to maximize the correct decision probability $%
P_{c}~$and minimize the error probability $P_{e}=1-P_{c}$.

We consider quantum state discrimination \cite{Eldar etal,Assalini
etal,Helstrom,Bergou} of the states in our system. For
example, we can use Helstrom's quantum detection and estimation theory \cite%
{Helstrom}, see \cite{Bergou} for a review and references therein. We first
consider the binary system of two pure coherent states $\rho (\Lambda
_{(0)}) $ and $\rho (\Lambda _{(1)})$ of the first type in Sec. \ref{sec 2.1}%
, and then the binary system of $|\Lambda _{(0)}\rangle _{k}$ and$~|\Lambda
_{(1)}\rangle _{k}$ of the second type in Sec \ref{sec 5.2}.

The correct decision probability $P_{c}$ is \cite{Helstrom,Bergou}
\begin{eqnarray}
P_{c} &=&\eta _{0}\mathrm{Tr}[\Pi _{0}\rho _{0}]+\eta _{1}\mathrm{Tr}[\Pi
_{1}\rho _{1}]  \notag \\
&=&\eta _{0}+\mathrm{Tr}[\Pi _{1}(\eta _{1}\rho _{1}-\eta _{0}\rho
_{0})]=:\eta _{0}+\mathrm{Tr}[\Pi _{1}D],
\end{eqnarray}%
where $\Pi _{0},\Pi _{1}$ are projectors and $\eta _{0}+\eta _{1}=1$, $\Pi
_{0}+\Pi _{1}=I$. The error probability is complementary to the correct
decision probability and is hence
\begin{equation}
P_{e}=\eta _{1}\mathrm{Tr}[\Pi _{0}\rho _{1}]+\eta _{0}\mathrm{Tr}[\Pi
_{1}\rho _{0}]=1-P_{c}.
\end{equation}

We maximize $P_{c}$,%
\begin{equation}
\max ~P_{c}=\eta _{0}+\max_{\{\Pi _{0},~\Pi _{1}\}}~\mathrm{Tr}[\Pi _{1}D].
\end{equation}%
The $D$ has eigen-decomposition%
\begin{equation}
D=\eta _{1}\rho _{1}-\eta _{0}\rho _{0}=\sum_{\lambda _{m}>0}\lambda
_{m}|\lambda _{m}\rangle \langle \lambda _{m}|+\sum_{\lambda _{m}<0}\lambda
_{m}|\lambda _{m}\rangle \langle \lambda _{m}|.
\end{equation}%
To maximize$~P_{c}$, we have $\Pi _{1}=\sum_{\lambda _{m}>0}|\lambda
_{m}\rangle \langle \lambda _{m}|$ and $\mathrm{Tr}[\Pi _{1}D]=\sum_{\lambda
_{m}>0}\lambda _{m}$, and hence
\begin{equation}
\max ~P_{c}=\eta _{0}+\sum_{\lambda _{m}>0}\lambda _{m}.
\end{equation}%
In the case there are multiple degenerate positive eigenvalues, all these
degenerate positive eigenvalues should be added.

Here, $\rho _{0}=\rho (\Lambda _{(0)})~$and $\rho _{1}=\rho (\Lambda _{(1)})$
are two pure multi-mode coherent states, as (\ref{density matrix Coh}). The
inner product of these two pure coherent states is%
\begin{eqnarray}
&&\frac{1}{\sqrt{\mathcal{N}(\Lambda _{(0)})\mathcal{N}(\Lambda _{(1)})}}%
\langle Coh(\Lambda _{(0)})|Coh(\Lambda _{(1)})\rangle  \notag \\
&&=\frac{\left( 1-|\Lambda _{(0)}|^{2}\right) ^{1/2}\left( 1-|\Lambda
_{(1)}|^{2}\right) ^{1/2}}{(1-\bar{\Lambda}_{(0)}\Lambda _{(1)})}.
\end{eqnarray}%
The eigenvalues of $D=\eta _{1}\rho (\Lambda _{(1)})-\eta _{0}\rho (\Lambda
_{(0)})$ are%
\begin{equation}
\lambda _{\pm }=\frac{1}{2}(\eta _{1}-\eta _{0}\pm \sqrt{1-4\eta _{0}\eta
_{1}\mathcal{N}_{(0)}^{-1}\mathcal{N}_{(1)}^{-1}\vert \langle Coh(\Lambda
_{(0)})|Coh(\Lambda _{(1)})\rangle \vert ^{2}}).
\end{equation}%
Hence,
\begin{eqnarray}
\max ~P_{c} &=&\eta _{0}+\lambda _{+}  \notag \\
&=&\frac{1}{2}+\frac{1}{2}\sqrt{1-4\eta _{0}\eta _{1}\frac{\left( 1-|\Lambda
_{(0)}|^{2}\right) \left( 1-|\Lambda _{(1)}|^{2}\right) }{|1-\bar{\Lambda}%
_{(0)}\Lambda _{(1)}|^{2}}}.
\end{eqnarray}%
This is also Helstrom's bound.

In the special case $\Lambda _{(1)}=-\Lambda _{(0)}\neq 0,~$if~$\eta
_{0}=\eta _{1}=\frac{1}{2}$,%
\begin{equation}
\max ~P_{c}=\frac{1}{2}+\frac{|\Lambda _{(1)}|}{1+|\Lambda _{(1)}|^{2}}%
,~~~~~~~\min ~P_{e}=\frac{1}{2}-\frac{|\Lambda _{(1)}|}{1+|\Lambda
_{(1)}|^{2}}.
\end{equation}%
In the special case $\Lambda _{(0)}=0,\Lambda _{(1)}\neq 0$,$~$and hence%
\begin{eqnarray}
\max ~P_{c} &=&\frac{1}{2}+\frac{1}{2}\sqrt{1-4\eta _{0}\eta _{1}\left(
1-|\Lambda _{(1)}|^{2}\right) } \\
&=&\frac{1}{2}+\frac{1}{2}|\Lambda _{(1)}|~\leq 1,~~~~\text{\textrm{if}~}%
\eta _{0}=\eta _{1}=\frac{1}{2},
\end{eqnarray}%
and in this case%
\begin{equation}
\min ~P_{e}=\frac{1}{2}-\frac{1}{2}|\Lambda _{(1)}|.~
\end{equation}%
In both above cases, the quantum error probability $P_{e}$ can be decreased
to zero, when $\Lambda _{(1)}=0$.

Similarly we can use two blobs $|\Lambda _{(0)}\rangle _{k},|\Lambda
_{(1)}\rangle _{k}~$centered at two different locations in the phase space,
and we have%
\begin{eqnarray}
\max ~P_{c} &=&\eta _{0}+\lambda _{+}  \notag \\
&=&\frac{1}{2}(1+\sqrt{1-4\eta _{0}\eta _{1}\vert \langle \Lambda
_{(0)}|\Lambda _{(1)}\rangle _{k}\vert ^{2}})  \notag \\
&=&\frac{1}{2}+\frac{1}{2}\sqrt{1-4\eta _{0}\eta _{1}e^{-\frac{1}{k}|\Lambda
_{(0)}-\Lambda _{(1)}|^{2}}}.
\end{eqnarray}%
If the two are far apart, the maximal probability can be very close to 1.

This method can be useful for discriminating the complicated microstates in
the quantum systems discussed in this paper, and also those complicated
microstates of fuzzballs. The distinguishing between different microstates
of fuzzballs is highly important for understanding the information of the
microstates and the total entropy. The ideas in quantum measurement theory
would be useful for this purpose. Hence we consider quantum detection in
this section. The quantum detection theory is particularly needed and
powerful when the states to distinguish in the measurement are
non-orthogonal.

\section{Entanglement and correlation in Hilbert spaces}

\label{sec 7} \renewcommand{\theequation}{7.\arabic{equation}} %
\setcounter{equation}{0} \renewcommand{\thethm}{7.\arabic{thm}} %
\setcounter{thm}{0} \renewcommand{\theprop}{7.\arabic{prop}} %
\setcounter{prop}{0}

\subsection{Entanglement and correlation in phase space in single Hilbert
space}

\label{sec 7.1}

Quantum states with entanglement and correlation in different regions of
phase space are very common, see e.g. the general and formal discussions
\cite{Almeida}. Now we want to make correlation and entanglement between
different regions in the phase space. We could have multi droplets and multi
rings in the phase space plane. In this section, we first consider two
entangled rings in single copy of Hilbert space and their correlation.

We consider entangled state of rings in single copy of Hilbert space
\begin{equation}
|\Psi _{\pm }\rangle =\frac{1}{\sqrt{2}}(|\Delta _{n_{1},k_{1}}\rangle
|\Delta _{n_{2},k_{2}}\rangle \pm |\Delta _{n_{2},k_{1}}\rangle |\Delta
_{n_{1},k_{2}}\rangle ).
\end{equation}%
This state is in $\mathcal{H}_{k_{1}}\otimes \mathcal{H}_{k_{2}}$. The first
term means that there is $k_{1}$ dual giants at location $n_{1}$ and $k_{2}$
dual giants at location $n_{2}$. The second term means that there is $k_{1}$
dual giants at location $n_{2}$ and $k_{2}$ dual giants at location $n_{1}$.
From the quantum measurement point of view, this is an entangled state
between a first set of $k_{1}$ dual giants$~$and a second set of $k_{2}$
dual giants. The Young tableau states describe both the giant graviton
states and the multi black ring states which are geometric spacetime
backgrounds. The multi black ring geometries are in terms of multi black
rings in the phase space plane. We make an approximation that the radii of
the two black rings are much larger than their widths. The areas of the
black rings are $k_{i}$.~The locations of the black rings are centered at $%
|z_{i}|^{2}=n_{i}+r_{0}^{2}$.~For two nearby rings labeled by $i$ and $%
i^{\prime }$, we have $|z_{i}|^{2}-|z_{i^{\prime }}|^{2}=n_{i}-n_{i^{\prime
}}$.$~$So this state is also an entangled state of two black rings centered
at $|z_{1}|$ and $|z_{2}|$ respectively in the phase space plane.

We consider the Hilbert space of the two rings to be $\mathcal{H}_{A}\otimes
\mathcal{H}_{B}=\mathcal{H}_{S}$, where $\mathcal{H}_{A}=\mathcal{H}_{k_{1}},%
\mathcal{H}_{B}=\mathcal{H}_{k_{2}}$. Consider the density matrix of this
entangled state $\rho _{S}$ in $\mathcal{H}_{A}\otimes \mathcal{H}_{B}$,
\begin{eqnarray}
\rho _{S} &=&\vert \Psi _{\pm }\rangle \langle \Psi _{\pm }\vert . \\
\rho _{A} &=&\mathrm{tr}_{\mathcal{H}_{B}}~\rho _{S}=\frac{1}{2}|\Delta
_{n_{1},k_{1}}\rangle \langle \Delta _{n_{1},k_{1}}|+\frac{1}{2}|\Delta
_{n_{2},k_{1}}\rangle \langle \Delta _{n_{2},k_{1}}|. \\
\rho _{B} &=&\mathrm{tr}_{\mathcal{H}_{A}}~\rho _{S}=\frac{1}{2}|\Delta
_{n_{1},k_{2}}\rangle \langle \Delta _{n_{1},k_{2}}|+\frac{1}{2}|\Delta
_{n_{2},k_{2}}\rangle \langle \Delta _{n_{2},k_{2}}|.
\end{eqnarray}

The correlation between $A$ and $B$ is
\begin{eqnarray}
C_{\rho _{S}}(\mathcal{O}_{A},\mathcal{O}_{B}):&=&\mathrm{tr~}(\rho _{S}%
\mathcal{O}_{A}\mathcal{O}_{B})-\mathrm{tr~}(\rho _{A}\mathcal{O}_{A})%
\mathrm{tr~}(\rho _{B}\mathcal{O}_{B})  \notag \\
&=&\mathrm{tr~}((\rho _{S}-\rho _{A}\otimes \rho _{B})\mathcal{O}_{A}%
\mathcal{O}_{B}),
\end{eqnarray}%
where $\mathcal{O}_{A}$ acts on $\mathcal{H}_{A}$, $\mathcal{O}_{B}$ acts on
$\mathcal{H}_{B}$,$~$and $\mathcal{O}_{A}\mathcal{O}_{B}$ acts on $\mathcal{H%
}_{A}\otimes \mathcal{H}_{B}.$

We have
\begin{eqnarray}
&&\mathrm{tr~}((\rho _{S}-\rho _{A}\otimes \rho _{B})\mathcal{O}_{A}\mathcal{%
O}_{B})  \notag \\
&=&\frac{1}{4}(\langle \mathcal{O}_{A}\rangle _{|\Delta
_{n_{1},k_{1}}\rangle }-\langle \mathcal{O}_{A}\rangle _{|\Delta
_{n_{2},k_{1}}\rangle })(\langle \mathcal{O}_{B}\rangle _{|\Delta
_{n_{1},k_{2}}\rangle }-\langle \mathcal{O}_{B}\rangle _{|\Delta
_{n_{2},k_{2}}\rangle })  \notag \\
&&\pm \frac{1}{2}[\langle \Delta _{n_{1},k_{1}}|\mathcal{O}_{A}|\Delta
_{n_{2},k_{1}}\rangle \langle \Delta _{n_{2},k_{2}}|\mathcal{O}_{B}|\Delta
_{n_{1},k_{2}}\rangle +\langle \Delta _{n_{2},k_{1}}|\mathcal{O}_{A}|\Delta
_{n_{1},k_{1}}\rangle \langle \Delta _{n_{1},k_{2}}|\mathcal{O}_{B}|\Delta
_{n_{2},k_{2}}\rangle ].  \notag \\
&&
\end{eqnarray}%
In the last line, the sign is $\pm $ for $\vert \Psi _{\pm }\rangle $
respectively.

With $\mathcal{O}_{A}=\hat{E}=\sum\limits_{k}A_{k}^{\dagger }A_{k},\mathcal{O%
}_{B}=\hat{E}=\sum\limits_{k}A_{k}^{\dagger }A_{k},$ we have%
\begin{eqnarray}
C_{\rho _{S}}(\hat{E}_{A},\hat{E}_{B}) &=&\mathrm{tr~}((\rho _{S}-\rho
_{A}\otimes \rho _{B})\hat{E}_{A}\hat{E}_{B})  \notag \\
&=&\frac{1}{4}(\langle \hat{E}_{A}\rangle _{|\Delta _{n_{1},k_{1}}\rangle
}-\langle \hat{E}_{A}\rangle _{|\Delta _{n_{2},k_{1}}\rangle })(\langle \hat{%
E}_{A}\rangle _{|\Delta _{n_{1},k_{2}}\rangle }-\langle \hat{E}_{A}\rangle
_{|\Delta _{n_{2},k_{2}}\rangle })  \notag \\
&=&\frac{1}{4}k_{1}k_{2}(n_{1}-n_{2})^{2}  \notag \\
&=&\frac{1}{4}k_{1}k_{2}(|z_{1}|^{2}-|z_{2}|^{2})^{2}.
\end{eqnarray}%
In the above, $\langle \hat{E}\rangle _{|\Delta _{n_{i},k_{i}}\rangle
}=k_{i}n_{i}$.

The area, i.e. $\mathrm{area(annulus)}$,\textrm{~}between the two black
rings is~proportional to $|z_{1}|^{2}-|z_{2}|^{2}=n_{1}-n_{2}=\frac{1}{2\pi
\hbar }~\mathrm{area(annulus)}$ \cite{Lin:2004nb}. Hence
\begin{eqnarray}
C_{\rho _{S}}(\hat{E}_{A},\hat{E}_{B}) &=&\mathrm{tr~}((\rho _{S}-\rho
_{A}\otimes \rho _{B})\hat{E}_{A}\hat{E}_{B})  \notag \\
&=&\frac{1}{16\pi ^{2}\hbar ^{2}}k_{1}k_{2}~\mathrm{area(annulus)}^{2}.
\end{eqnarray}%
The amount of correlation is proportional to the area squared of the annulus
between the two rings. This is a geometric interpretation of the correlation
and entanglement between the two rings. This annulus configuration can be
viewed as a `bridge' between the two rings, in which the two rings are
entangled. Similar scenario has also been pointed out in \cite{Simon:2018laf,Lin:2017dnz}.

\subsection{States in doubled Hilbert space and purification}

\renewcommand{\theequation}{8.\arabic{equation}} \setcounter{equation}{0} %
\renewcommand{\thethm}{8.\arabic{thm}} \setcounter{thm}{0} %
\renewcommand{\theprop}{8.\arabic{prop}} \setcounter{prop}{0}

Now we want to make correlation or entanglement between two copies of the
phase space, and further relate this to the connectivity between the two
copies of the spacetime \cite{Simon:2018laf}. This is different from the
preceding Sec \ref{sec 7.1} for single copy. The doubled Hilbert space has
been considered in \cite{Simon:2018laf}. We are interested in gray rings and
doubled Hilbert space, since the gray rings carry entropy and one can
perform a purification of the gray ring states to a pure state in doubled
Hilbert space.

The $|\Psi _{1}\rangle ,|\Psi _{2}\rangle $ are pure states in single copy
of Hilbert space:%
\begin{eqnarray}
|\Psi _{1}\rangle &=&\sqrt{\frac{k_{1}}{N}}|\Delta _{n_{1},k_{1}}\rangle +%
\sqrt{\frac{k_{2}}{N}}|\Delta _{n_{2},k_{2}}\rangle , \\
|\Psi _{2}\rangle &=&\sqrt{\frac{k_{1}}{N}}|\Delta _{n_{1},k_{1}}\rangle +%
\sqrt{\frac{k_{2}}{N}}|\Delta _{n_{3},k_{2}}\rangle ,  \label{YT sup}
\end{eqnarray}%
where $k_{1}+k_{2}=N$, and $\vert \Psi _{1}\vert =\vert \Psi _{2}\vert =1$.
We make an approximation that\ $|\Psi _{1}\rangle $ and $|\Psi _{2}\rangle $
are approximately orthogonal, and this means that $k_{1}\ll N$. The YT has
better orthogonality than the coherent state, and as a result the
separations between $n_{1},n_{2},n_{3}$ needn't be large. This gives the
pure state in doubled system%
\begin{equation}
\vert \Psi \rangle =\alpha \vert \Psi _{1}\rangle \vert \Psi _{1}\rangle +%
\sqrt{1-\alpha ^{2}}\vert \Psi _{2}\rangle \vert \Psi _{2}\rangle
\end{equation}%
which is in $\mathcal{H}^{L}\otimes \mathcal{H}^{R}$.\ We have $\vert \Psi
\vert =1$. The density matrix $\rho ~$of the doubled system is%
\begin{equation}
\rho =\vert \Psi \rangle \langle \Psi \vert .  \label{density matrix 06}
\end{equation}

The density matrix after partial tracing from the above density matrix (\ref%
{density matrix 06}) is%
\begin{equation}
\rho _{L}=\mathrm{tr}_{\mathcal{H}^{R}}~\vert \Psi \rangle \langle \Psi
\vert \simeq \alpha ^{2}\vert \Psi _{1}\rangle \langle \Psi _{1}\vert
+(1-\alpha ^{2})\vert \Psi _{2}\rangle \langle \Psi _{2}\vert ,
\end{equation}%
which is derived by assuming the orthogonality between $|\Psi _{1}\rangle $
and $|\Psi _{2}\rangle $. And similarly we have $\rho _{R}=\mathrm{tr}_{%
\mathcal{H}^{L}}~\vert \Psi \rangle \langle \Psi \vert $. The $\rho $ can be
viewed as a purification of $\rho _{L}$. Hence,
\begin{eqnarray}
\rho _{L} &=&\mathrm{tr}_{\mathcal{H}^{R}}~\vert \Psi \rangle \langle \Psi
\vert  \notag \\
&\simeq &\frac{k_{1}}{N}|\Delta _{n_{1},k_{1}}\rangle \langle \Delta
_{n_{1},k_{1}}|+\frac{k_{2}}{N}\alpha ^{2}|\Delta _{n_{2},k_{2}}\rangle
\langle \Delta _{n_{2},k_{2}}|+\frac{k_{2}}{N}(1-\alpha ^{2})|\Delta
_{n_{3},k_{2}}\rangle \langle \Delta _{n_{3},k_{2}}|  \notag \\
&&+\sqrt{\frac{k_{1}k_{2}}{N^{2}}}(\alpha ^{2}|\Delta _{n_{2},k_{2}}\rangle
+(1-\alpha ^{2})|\Delta _{n_{3},k_{2}}\rangle )\langle \Delta
_{n_{1},k_{1}}|~+\mathrm{h.c.}
\end{eqnarray}%
Here $k_{1}+k_{2}=N$ and $\mathrm{tr}\rho _{L}=1$. This state corresponds to
a black droplet with filling fraction 1 and two gray rings located around $%
|z_{2}|$ and $|z_{3}|$ respectively, with filling fraction which is not 1,
but $\alpha ^{2}$ and $(1-\alpha ^{2})$ respectively. And we have $%
|z_{2}|^{2}=n_{2}+r_{0}^{2}~$and $|z_{3}|^{2}=n_{3}+r_{0}^{2}$. This is a
superposed mixed state. We make an approximation that the radii of the two
rings are much larger than their widths. Note that the locations of the
rings are centered at $|z_{i}|^{2}=n_{i}+r_{0}^{2}$, and $%
|z_{i}|^{2}-|z_{i^{\prime }}|^{2}=n_{i}-n_{i^{\prime }}$ for two nearby
rings.

The correlation between left and right systems is:
\begin{eqnarray}
C_{\rho }(\mathcal{O}_{L},\mathcal{O}_{R}):&=&\mathrm{tr~}(\rho \mathcal{O}%
_{L}\mathcal{O}_{R})-\mathrm{tr~}(\rho _{L}\mathcal{O}_{L})\mathrm{tr~}(\rho
_{R}\mathcal{O}_{R})  \notag \\
&=&\mathrm{tr~}((\rho -\rho _{L}\otimes \rho _{R})\mathcal{O}_{L}\mathcal{O}%
_{R}),
\end{eqnarray}%
where $\mathcal{O}_{L}$ acts on $\mathcal{H}^{L}$, $\mathcal{O}_{R}$ acts on
$\mathcal{H}^{R}$,$~$and $\mathcal{O}_{L}\mathcal{O}_{R}$ acts on $\mathcal{H%
}^{L}\otimes \mathcal{H}^{R}=\mathcal{H}^{S}$. We denote the total system as
$S=L\cup R$, and $\rho =\rho _{S}.$

We can have $\mathcal{O}_{L}:=\mathcal{O}=\exp (i\theta \hat{E})=\exp
(i\theta \sum\limits_{k}\hat{E}_{k})$, where $\hat{E}_{k}=A_{k}^{\dagger
}A_{k}$. And we have $\hat{E}_{L}=\sum\limits_{k}\hat{E}_{k;L}=\sum%
\limits_{k}A_{k;L}^{\dagger }A_{k;L}$,~$\hat{E}_{R}=\sum\limits_{k}\hat{E}%
_{k;R}=\sum\limits_{k}A_{k;R}^{\dagger }A_{k;R}$. We define%
\begin{equation}
F_{\theta _{L},\theta _{R}}=\mathrm{tr~}((\rho -\rho _{L}\otimes \rho
_{R})\exp (i\theta _{L}\hat{E}_{L})\exp (i\theta _{R}\hat{E}_{R})).
\end{equation}%
We can calculate $\mathcal{O}_{L}:=\mathcal{O}$, where $\mathcal{O}=\hat{E}%
=\sum\limits_{k}\hat{E}_{k}=\sum\limits_{k}A_{k}^{\dagger }A_{k}$, and
\begin{equation}
\mathrm{tr~}((\rho -\rho _{L}\otimes \rho _{R})\hat{E}_{L}\hat{E}%
_{R})=-\partial _{\theta _{L}}\partial _{\theta _{R}}F_{\theta _{L},\theta
_{R}}|_{\theta _{L}=\theta _{R}=0}.
\end{equation}

With the density matrix above, for a general $\mathcal{O}$,
\begin{eqnarray}
&&\mathrm{tr~}((\rho -\rho _{L}\otimes \rho _{R})\mathcal{O}_{L}\mathcal{O}%
_{R})  \notag \\
&\simeq &\alpha ^{2}(1-\alpha ^{2})[(\langle \mathcal{O}_{L}\rangle _{|\Psi
_{1}\rangle }-\langle \mathcal{O}_{L}\rangle _{|\Psi _{2}\rangle })(\langle
\mathcal{O}_{R}\rangle _{|\Psi _{1}\rangle }-\langle \mathcal{O}_{R}\rangle
_{|\Psi _{2}\rangle })]  \notag \\
&&+\alpha \sqrt{1-\alpha ^{2}}[\langle \Psi _{1}|\mathcal{O}_{L}|\Psi
_{2}\rangle \langle \Psi _{1}|\mathcal{O}_{R}|\Psi _{2}\rangle +\langle \Psi
_{2}|\mathcal{O}_{L}|\Psi _{1}\rangle \langle \Psi _{2}|\mathcal{O}_{R}|\Psi
_{1}\rangle ].
\end{eqnarray}%
This is a special case of the more general formalism in \cite{Simon:2018laf}.

For $\mathcal{O}=\exp (i\theta \hat{E})$, we have that
\begin{eqnarray}
&&\mathrm{tr~}((\rho -\rho _{L}\otimes \rho _{R})\exp (i\theta _{L}\hat{E}%
_{L})\exp (i\theta _{R}\hat{E}_{R}))  \notag \\
&\simeq &\alpha ^{2}(1-\alpha ^{2})\frac{k_{2}^{~2}}{N^{2}}%
[(e^{ik_{2}n_{2}\theta _{L}}-e^{ik_{2}n_{3}\theta
_{L}})(e^{ik_{2}n_{2}\theta _{R}}-e^{ik_{2}n_{3}\theta _{R}})].
\end{eqnarray}%
We denote $S=L\cup R$ and
\begin{eqnarray}
\langle \hat{E}_{S}\rangle_{\rho } &=& \langle \hat{E}_{L}+\hat{E}%
_{R}\rangle _{\rho }=\langle \hat{E}_{L}\rangle _{\rho }+\langle \hat{E}%
_{R}\rangle _{\rho }=\langle \hat{E}_{L}\rangle _{\rho _{L}}+\langle \hat{E}%
_{R}\rangle _{\rho _{R}},  \notag \\
\langle \hat{E}_{L}\rangle_{\rho } &=& \langle \hat{E}_{R}\rangle _{\rho
}~\simeq ~\frac{k_{1}^{~2}}{N}n_{1}+\frac{k_{2}^{~2}}{N}(\alpha
^{2}n_{2}+(1-\alpha ^{2})n_{3}).
\end{eqnarray}%
On the YT states, we have $\langle \hat{E}_{L}\rangle _{|\Psi _{i}\rangle }=%
\frac{k_{1}^{~2}}{N}n_{1}+\frac{k_{2}^{~2}}{N}n_{i+1}$, $\langle \hat{E}%
_{L}\rangle _{|\Psi _{i}\rangle }-\langle \hat{E}_{L}\rangle _{|\Psi
_{i^{\prime }}\rangle }=\frac{k_{2}^{~2}}{N}(n_{i+1}^{~}-n_{i^{\prime
}+1}^{~})$.

We hence have%
\begin{eqnarray}
\mathrm{tr~}((\rho -\rho _{L}\otimes \rho _{R})\hat{E}_{L}\hat{E}_{R})
&=&-\partial _{\theta _{L}}\partial _{\theta _{R}}F_{\theta _{L},\theta
_{R}}|_{\theta _{L}=\theta _{R}=0}  \notag \\
&\simeq &\alpha ^{2}(1-\alpha ^{2})\frac{k_{2}^{~2}}{N^{2}}%
k_{2}^{~2}(n_{2}-n_{3})^{2}.
\end{eqnarray}%
This is always non-zero as long as $n_{2}\neq n_{3}$. The locations of the
rings are centered at $|z_{i}|^{2}$, and $|z_{i}|^{2}-|z_{i^{\prime
}}|^{2}=n_{i}-n_{i^{\prime }}$.

Since $n_{2}-n_{3}=|z_{2}|^{2}-|z_{3}|^{2}=\frac{1}{2\pi \hbar }~\mathrm{%
area(annulus)}$ \cite{Lin:2004nb}, the correlation between the left and
right is
\begin{eqnarray}
C_{\rho }(\hat{E}_{L},\hat{E}_{R}) &=&\mathrm{tr~}((\rho -\rho _{L}\otimes
\rho _{R})\hat{E}_{L}\hat{E}_{R})  \notag \\
&\simeq &\frac{1}{4\pi ^{2}\hbar ^{2}}\alpha ^{2}(1-\alpha ^{2})\frac{%
k_{2}^{~4}}{N^{2}}~\mathrm{area(annulus)}^{2}.
\end{eqnarray}%
It is related to the area of annulus between two gray rings.\ There is a
square because we have the product of two operators.

This is related to the superstar spacetime \cite{Myers:2001aq,Balasubramanian:2005mg,DErrico:2007qch}. The microstates include
gray rings whose filling fraction is smaller than 1. The gray rings are the
analog of black hole horizon, in the sense they encode information of the
microstates and carry entropy.

These density matrices are similar to that of the superstar geometry. We
have to sum over the gray rings and hence in the geometric dual the gray
rings must be glued so that they disappear after the gluing. After the
gluing, since there is no gray droplets left, it becomes a pure state in the
doubled system. This supports the proposal that the doubled system is glued
along the gray droplets \cite{Simon:2018laf}. This indicates that the above
doubled system is glued along the two gray rings. The `bridge' or `tunnel'
configuration may be viewed as the annulus between the two gray rings \cite%
{Simon:2018laf}.

\section{Discussion}

\label{sec_discussion}

In this paper, we focused on two types of coherent states. One type \cite%
{Berenstein:2017abm} is the coherent states of multi traces, and the other
type is the coherent states of multi columns. The first type can be viewed
as the coherent states of multi gravitons. In the second type, each column
can be viewed as a giant graviton or a D-brane. Hence, the second type can
be viewed as the coherent states of giant gravitons.

Among other things, we also analyzed interesting states in the phase space,
the bumps on the edge of droplet, blobs and rings. The bumps are coherent
states of multi-traces. While the rings are Young tableau states. Adding the
bumps does not change the topology of the droplet configuration, while
adding blobs or rings changes the topology.

We can use more matrices \cite{Lin:2017vfn,Balasubramanian:2018yjq}
to construct the coherent states of these two types for the multi graviton
states and giant graviton states. They have a rich entanglement structure.
With more matrices, we can construct more unitary operations generalizing
the phase shift operation, since different types of matrices can have
independent phase rotations.

The gauge/gravity duality enables us to analyze the aspects of superposition
and entanglement for the quantum gravity side. The gravitational aspects of
the gravitational superposition states have also been discussed in \cite%
{Berenstein:2017abm,Berenstein:2016pcx,Berenstein:2016mxt,Anastopoulos:2015zta}. The cat states in Newtonian gravity and the
associated fluctuations of gravitational potential and its effect on the
probes in the spacetime background have been discussed in \cite%
{Anastopoulos:2015zta}. These probes have time-dependent dynamics. We can
also study various probes on the backgrounds similar to the ones related to
this paper, such as multi gravitons, closed strings, giant magnons and probe
branes.

We found $N$-state Schrodinger cat states which approach the one-row Young
tableau states with fidelity between them asymptotically reaches 1 at large $%
N$. This is of significance that angularly localized states superpose to
form angularly delocalized state in quantum gravity. This superposition cat
state has high variance of excitation energy. The $N$-state Schrodinger cat
states also correspond to the irreducible representations of the cyclic
group.

We conveniently used a phase shift operator, which also provides a
generating function for the excitation energy operator. The phase shift
operator also gives an associated parity operator which can be used to
define cat states. The quantum Fisher information of these states is
proportional to the variance of the excitation energy of the underlying
states. Some states like the first type of coherent states have high quantum
Fisher information, and as a result have good localizability in the angular
direction in the phase space. Some other states have relatively low quantum
Fisher information and as a result are delocalized in the angular direction
in the phase space.

We also analyzed correlation and entanglement between gravitational degrees
of freedom using the phase space. In the context of single sided Hilbert
space, the correlation between two entangled ring shaped droplets in phase
space is related to the area of the annulus between the two rings. In the
context of the doubled Hilbert space, the correlation between two sets of
rings entangled in the doubled Hilbert space is related to the area of the
annulus between the two rings, where the annulus plays the role of the
`bridge'.

The approach of correlation and entanglement in phase space \cite{Almeida}
is convenient for studying questions with gravitational degrees of freedom
using different regions of the phase space. The setup in this context
provides a laboratory for studying these quantum gravitational questions.

In the context of gauge/gravity duality, it has been observed that the
entanglement between different parts of the bulk spacetime or universe is
related to the connectivity between the different parts \cite%
{Maldacena:2013xja,VanRaamsdonk:2010pw,Simon:2018laf,Lin:2017dnz}. Our findings are in agreement with this scenario.

We considered two types of noisy coherent states and their derivations from
adding noise. These are generalizations of the noisy coherent states of
photons. The noisy coherent state can be viewed as an interpolated state. It
interpolates between a pure coherent state in the noiseless limit, and a
maximally mixed state in the large noise limit. The later has well-known
gravity dual interpretation as thermal gas or black holes. For the noisy
coherent state of the second type, the random columns are similar to a gas
of thermal giant gravitons. The thermal distributions play the role of the
noise. Moreover, we have a purification of the limit of the noisy coherent
states. The noisy coherent states contain both quantum information and
noise, and are widely used in quantum information theory.

The multitude of these different microstates in the quantum systems
discussed in this paper are similar to the microstates of fuzzballs \cite%
{Mathur:2005ai,Bena:2007kg,Skenderis:2008qn,Balasubramanian:2008da}, which have provided important insights into the
information paradox. The quantum detection and quantum measurement theory
are useful for distinguishing different quantum states of a quantum system.

\section*{Acknowledgments}

We would like to thank B.~Czech, R.~de Mello Koch and J. Simon for
communications or discussions. The work was supported in part by Yau
Mathematical Sciences Center and Tsinghua University, and by grant
TH-533310008 of Tsinghua University (to H.L.).

\end{document}